\documentclass[sigconf]{acmart}
\AtBeginDocument{%
  }

\setcopyright{rightsretained}
\copyrightyear{2026}
\acmYear{2026}
\acmDOI{XXXXXXX.XXXXXXX}

\acmConference[]{}{}{}


\newcommand{\myparagraph}[1]{ \noindent{\bf {#1}.}}

\newcommand{\out}[1] {}

\newcounter{codeLineCntr}

\reversemarginpar
\newif\ifnotes
\notestrue

\newcommand{\punt}[1]{}

\renewcommand{\eqref}[1]{Equation~(\ref{eq:#1})}

\newcommand{\proc}[1]{\ifmmode\mbox{\textsc{#1}}\else\textsc{#1}\fi}

\newcommand{\func}[1]{\ifmmode\mathrm{#1}\else\textrm{#1}fi} %

\newcounter{remark}[section]

\usepackage{ifthen}

\newcommand{\MP}[1]{\textcolor{red}{#1}}

\newboolean{publicversion}
\setboolean{publicversion}{false}

\ifthenelse{\boolean{publicversion}}{
   \newcommand{\grumbler}[2]{}
   
}{
   \newcommand{\grumbler}[2]{\MP{{\bf #1}: #2}}
    
}

\newcommand{\dimitris}[1]{\grumbler{Dimitris}{#1}}
\newcommand{\masa}[1]{\grumbler{Masa}{#1}}

\newboolean{appendix}
\setboolean{appendix}{true}

\usepackage[small,compact,noindentafter]{titlesec} 
\usepackage[compact]{titlesec}

\titlespacing\section{0pt}{6pt plus 2pt minus 2pt}{2pt plus 2pt minus 2pt}
\titlespacing\subsection{0pt}{4pt plus 2pt minus 2pt}{2pt plus 1pt minus 1pt}
\titlespacing{\paragraph}{0pt}{2pt plus 0pt minus 1pt}{1.0ex}

\usepackage[inline]{enumitem}
\setlist{noitemsep,topsep=0pt,parsep=0pt,partopsep=0pt}

\usepackage[subtle,margins=normal]{savetrees}

\usepackage{setspace}
\usepackage[margin=5pt,font={stretch=0.9}]{caption}

\usepackage{tikz}
\usepackage{amsmath}
\usepackage{xspace}
\usepackage{xcolor}
\usepackage{pifont}%
\usepackage{enumitem}
\usepackage{multirow}
\usepackage{algorithmic}
\usepackage[ruled,vlined,linesnumbered]{algorithm2e}

\usepackage{hyperref}
\usepackage{tcolorbox}
\tcbuselibrary{skins} %
\usepackage{multicol}

\usepackage{cuted}

\usepackage{subcaption}
\usepackage{makecell}
\usepackage{circledsteps}

\usepackage{listings}
\definecolor{codegreen}{rgb}{0,0.6,0}
\definecolor{codegray}{rgb}{0.5,0.5,0.5}
\definecolor{codepurple}{rgb}{0.58,0,0.82}
\definecolor{backcolour}{rgb}{0.95,0.95,0.92}
\lstdefinestyle{customc}{
   backgroundcolor=\color{backcolour},   
   commentstyle=\color{codegreen},
   keywordstyle=\color{magenta},
   numberstyle=\tiny\color{codegray},
   stringstyle=\color{codepurple},
   basicstyle=\ttfamily\footnotesize\linespread{0.7},
   breaklines,
   tabsize=2,
   numbers=left,
   columns=fullflexible,
   keepspaces=true,
   frame=lines,
   numbersep=1pt,
   escapechar=|,
   captionpos=b,
   language=c,
   keywords={query, sessionKey, objPurIs,time_threshold, objObjections,objOwnIs, objPur, objOrig, objShare, objExp, objObj, monitor, encryption},
   aboveskip=1pt,        %
   belowskip=1pt,        %
   framextopmargin=-2pt,  %
   framexbottommargin=-2pt, %
   framesep=1pt,
   rulesep=0pt,
}
\lstdefinestyle{embedded}{
 numbers=none,
 frame=none,
 xleftmargin=0cm,
 backgroundcolor=\color{Lavender},
 framesep=1pt,
 aboveskip=3pt,
 belowskip=3pt,
}

\newcommand{\projectname}{\textsc{GDPRuler}\xspace}
\newcommand{\gdpr}{\textsc{GDPR}\xspace}

\definecolor{darkpastelgreen}{rgb}{0.01, 0.75, 0.24}
\definecolor{darkpastelred}{rgb}{0.76, 0.23, 0.13}
\newcommand{\cmark}{\textcolor{darkpastelgreen}{\ding{51}}}%
\newcommand{\xmark}{\textcolor{darkpastelred}{\ding{55}}}%

\newcommand{\myfontsizealgorithms}{\fontsize{7}{7}\selectfont}

\setcopyright{none}
\renewcommand\footnotetextcopyrightpermission[1]{} %

\begin{document}

\date{}

\title{Policy-Compliant Cloud Storage Systems}

\author{
{\rm Dimitrios Stavrakakis}\textsuperscript{1} \quad {\rm Masanori Misono}\textsuperscript{1} \quad {\rm Julian Pritzi}\textsuperscript{1} \\ {\rm Harshavardhan Unnibhavi}\textsuperscript{1} \quad {\rm Nuno Santos}\textsuperscript{2} \quad {\rm Pramod Bhatotia}\textsuperscript{1}\\
\textsuperscript{1}Technical University of Munich \quad
\textsuperscript{2}INESC-ID/Instituto Superior Tecnico, University of Lisbon
}

\renewcommand{\shortauthors}{Dimitrios Stavrakakis et al.}

\begin{abstract}

Privacy regulations such as the General Data Protection Regulation (\gdpr) impose strict requirements on how personal data is stored, processed, and audited. 
While key-value stores (KVS) are widely used in latency-sensitive applications, their simple data model and untrusted cloud deployment environments make \gdpr compliance particularly challenging. 
Existing approaches require invasive code modifications, impose high performance overheads, or overlook the integrity of compliance mechanisms themselves.

This paper presents \projectname{}, a trusted middleware system that enables verifiable \gdpr compliance for KVS on untrusted clouds without modifying their codebase. 
\projectname{} deploys a \emph{trusted \gdpr monitor} inside a Confidential Virtual Machine (CVM), which enforces \gdpr policies, manages compliance metadata, and maintains tamper-evident audit logs. 
A declarative policy language translates core \gdpr obligations into enforceable runtime rules. 
To ensure efficiency, \projectname{} encodes metadata compactly within KV records, builds dedicated metadata indexes for \gdpr-specific queries, and logs only compliance-relevant events in a space-efficient format.
We implement \projectname{} as a transparent proxy for unmodified Redis and RocksDB deployments. Evaluation with YCSB and \gdpr-inspired workloads shows that \projectname{} enforces core compliance guarantees with low overheads: \projectname{} achieves $\approx$61\% of native KVS throughput with the CVM environment contributing 28\%-32\% of it, metadata storage overhead remains below 20\%, and \gdpr queries benefit from 13-182$\times$ speedup through metadata indexing. By embedding verifiable policy enforcement into a trusted middleware layer, \projectname{} offers a practical path toward \gdpr-compliant KVS on untrusted cloud infrastructures.

\end{abstract}

\maketitle
\pagestyle{plain}
\renewcommand{\sectionautorefname}{\S\!}
\renewcommand{\subsectionautorefname}{\S\!}
\renewcommand{\subsubsectionautorefname}{\S\!}

\section{Introduction}

The rapid move of data processing and storage to the cloud has shifted control of sensitive information from individuals and organizations to untrusted third‑party infrastructures, intensifying the concerns over data security, privacy, and anonymity.
Across diverse cloud data architectures---from data warehouses to distributed data processing systems---modern privacy regulations such as the General Data Protection Regulation (\gdpr)~\cite{regulation-gdpr} require data controllers to operationalize users’ rights---such as access and erasure~\cite{gdpr-art-17,gdpr-art-18}---and to uphold principles of purpose and storage limitation~\cite{gdpr-art-5} directly in their storage and processing pipelines. 
Non-compliance can result in severe financial and reputational penalties~\cite{noyb,enforcement-tracker-gdpr,google-gemini,meta-data-transfers}, motivating systems that can not only protect data but also \emph{record, enforce, and prove} compliance actions through auditable operations on stored data. 
Achieving this goal in cloud deployments, however, remains difficult: privileged software stacks controlled by cloud operators can observe or tamper with data and audit trails, undermining both confidentiality and the integrity of compliance evidence.

Key-Value Stores (KVS)~\cite{redis,rocksdb,kv-ranking,memcached,dynamo-db} constitute a core component of modern cloud data architectures, exemplifying the tension between simplicity and compliance. 
They underpin a wide range of modern data services, from caching layers to IoT pipelines, but their flat key-value (KV) abstraction provides no native support for compliance metadata, secondary indexes, or expressive queries. 
As a result, mechanisms such as purpose-based access must be implemented above the data layer, through ad hoc application logic that offers little assurance of correctness or completeness. 
Unlike relational databases, KVS lack schemas and query semantics that would allow them to represent and reason about regulatory state.

A fundamental design question is thus \emph{where and how to enforce privacy-aware compliance}. Embedding GDPR mechanisms directly into KV engines (e.g., Redis~\cite{redis}, RocksDB~\cite{rocksdb}, LevelDB~\cite{leveldb}) would tightly couple regulatory policy with system internals, complicating upgrades and breaking compatibility. Delegating enforcement to the application layer scatters compliance checks and forfeits verifiability.
We argue that effective \gdpr compliance requires a dedicated, verifiable \emph{trusted middleware layer}---a secure data-management service interposed between applications and unmodified KVS backends, which centralizes policy enforcement, metadata management, and audit logging while preserving the existing KVS API.

However, designing such a layer raises three intertwined challenges spanning \emph{security and privacy}, \emph{policy enforcement}, and \emph{performance}. First, the middleware must remain secure and trustworthy even when deployed on untrusted clouds, preventing privileged software from observing or altering compliance logic or logs. Administrators or compromised hypervisors may gain access to in-memory data, intercept compliance metadata, or tamper with audit logs to conceal non-compliant behavior~\cite{rocha_insider_2011}, undermining both confidentiality and the ability to demonstrate lawful processing. Existing approaches that rely on application-level controls~\cite{gdprizer} or data-engine enforcement~\cite{k9db,gdprbench} assume a trusted cloud stack and leave the integrity of compliance mechanisms (e.g., metadata) and audit logs unprotected.

Second, the system must translate legal requirements into enforceable policies that can be applied to KV operations without relying on schemas or rich query semantics. This translation is challenging because \gdpr requirements are written in legal rather than technical language, leaving system builders to decide how principles such as storage or purpose limitation should be represented and enforced. Prior work on compliance-aware relational databases~\cite{k9db,gdprbench} introduces heavy schema extensions and relies on trusted execution environments, but these designs do not generalize to lightweight, high-throughput KVS workloads or to untrusted cloud deployments.

Third, the middleware must preserve high performance while continuously enforcing compliance and maintaining observability. Integrating compliance logic in the data path risks undermining efficiency that makes KVS attractive. Associating metadata with every record increases storage footprint, while enforcing deletion and access rights often requires non-key lookups that degenerate into full scans. Prior analyses show that naïve \gdpr implementations can lead to high compliance-query latencies even for small datasets~\cite{gdprbench}. Moreover, generating tamper-evident audit logs introduces I/O and synchronization overheads that accumulate in high-throughput scenarios, exhausting resources and inflating latency.%

Our key insight is that \gdpr compliance can be realized as a \emph{verifiable data-management layer} that executes inside a Confidential Virtual Machine (CVM). 
Running the enforcement logic within this hardware-isolated environment allows the system to mediate all KV operations securely, associate each key with compact compliance metadata, and maintain cryptographically protected audit logs, while requiring no modifications to existing storage engines.

This paper presents \projectname{}, a trusted middleware that enables \gdpr-compliant KVS on the untrusted cloud. More precisely, it deploys a \emph{trusted compliance monitor} inside a CVM, providing hardware-backed isolation, integrity, and attestation guarantees for compliance logic and metadata. The monitor intercepts all KV operations, enforces \gdpr policies, maintains per-record metadata, and produces verifiable audit trails. To express compliance rules, \projectname{} introduces a declarative policy language that compiles legal requirements (e.g., purpose limitation) into lightweight runtime checks within the CVM. Compact metadata encoding and dedicated indexes accelerate compliance queries, while a tamper-evident logging system records compliance-relevant operations in a secure, space-efficient format. By acting as a transparent proxy, \projectname{} preserves compatibility with existing deployments, allowing organizations to retrofit \gdpr compliance without code changes.

We implement \projectname{} as a transparent proxy layer compatible with unmodified Redis and RocksDB deployments. 
Our evaluation using YCSB and \gdpr-inspired workloads demonstrates that \projectname{} enforces key \gdpr obligations while maintaining high throughput and modest storage overheads: \projectname{} achieves $\approx$61\% of native throughput on average, with the underlying CVM contributing 28-32\% of this overhead. \gdpr queries achieve 13-182$\times$ improvement through dedicated metadata indexing, while metadata storage overhead remains below 20\%, and the tamper-evident logging system reduces throughput by merely 2\%. 
These results confirm that \projectname{} delivers secure, policy-compliant, and auditable data management on untrusted clouds without sacrificing performance.

In summary, the main contributions of this paper are as follows:
\begin{itemize}[noitemsep,leftmargin=*]
  \item \textbf{Trusted middleware architecture:} a data-management layer that decouples compliance enforcement from applications and KV engines, providing compatibility and verifiable \gdpr control.
  \item \textbf{Secure enforcement environment:} a CVM-based \emph{trusted compliance monitor} that ensures confidentiality, integrity, and remote attestation on untrusted cloud infrastructures.
  \item \textbf{Declarative policy framework:} a policy language and compiler translating \gdpr rules to enforceable constraints in the data path.
  \item \textbf{Efficient metadata and auditing:} compact per-record metadata encoding, compliance-specific indexing, and tamper-evident logging optimized for high-throughput workloads.
  \item \textbf{Implementation and evaluation:} a prototype supporting Redis and RocksDB demonstrating secure, low-overhead \gdpr enforcement under realistic workloads.
\end{itemize}

\section{Background}
\subsection{Confidential Virtual Machines (CVMs)}
Confidential computing~\cite{confidential_computing} protects data in use through hardware-assisted Trusted Execution Environments (TEEs), while providing remote attestation capabilities to ensure the system is deployed with the intended configuration, enabling secure processing in untrusted cloud environments.
Recently, the industry has shifted toward VM-level confidential computing solutions, namely \emph{Confidential Virtual Machines (CVMs)}~\cite{tdx_survey,cvm-eval,cca-eval}, including AMD SEV-SNP~\cite{sev-snp}, Intel TDX~\cite{tdx}, and ARM CCA~\cite{cca}. Major cloud providers have started offering CVMs as a service~\cite{azure-confidential-computing,gcp-confidential-computing}. %
CVMs run the entire guest system within the TEE, enabling the reuse of existing applications 
while preventing unauthorized access even from the cloud provider.

\subsection{\gdpr Policy Compliance for KVS}
\label{sec:background}

\begin{figure}[t]
\centering
\includegraphics[width=0.85\columnwidth]{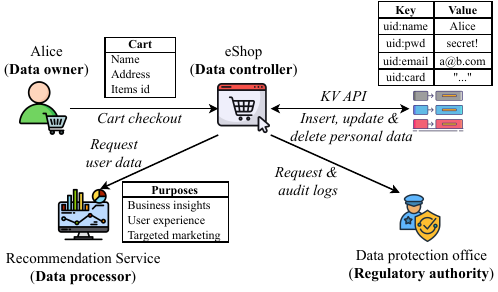}
\vspace{-2mm}
\caption{\gdpr ecosystem: {\em  The eShop (controller) stores user data in a KVS and uses a recommendation service (processor). Purpose limitation (Art. 5(1)(b)) restricts Alice's data usage to the declared purpose. The controller must provide verifiable compliance and respond to regulatory requests.}
}
\label{fig:gdpr-entities-colour}
\vspace{-1mm}
\end{figure}

KVS provide a simple data model and high performance, making them the backbone of latency-sensitive, large-scale services. 
At the same time, organizations are increasingly subject to strict regulatory frameworks that govern how personal data must be handled, including the General Data Protection Regulation (\gdpr) in Europe, the California Consumer Privacy Act (CCPA) in the United States, and other emerging data protection acts. 
Among these, \gdpr has become the global reference, imposing strict responsibilities and hefty penalties for non-compliance.
Since many KVS‑based applications manage personal data directly, ensuring \gdpr compliance is is both a legal and technical necessity. This paper focuses on providing system support for \emph{\gdpr-compliant KVS applications}, while preserving the efficiency and compatibility that make KVS attractive in practice.

\myparagraph{\gdpr entities in KVS applications} 
\gdpr defines distinct actors with specific obligations that naturally map to KVS workflows handling sensitive data (\autoref{fig:gdpr-entities-colour}). 
The \emph{data owner} is the individual whose personal data is stored in the KVS. 
The \emph{data controller} is the organization operating the application and bears the legal responsibility to ensure \gdpr compliance. 
The \emph{data processors} are internal or external services that perform operations on the KVS on behalf of the 
controller and must declare their intended purpose of data use. 
Finally, \emph{regulatory authorities} supervise compliance and may request verifiable logs to audit how personal data is processed.

\myparagraph{Compliance requirements for KVS}
To achieve \gdpr compliance all KV operations (e.g., \emph{get}/\emph{put}/\emph{delete}) must respect the policies defined by data owners, that processors’ declared purposes are checked against those policies, and that controllers maintain tamper-evident records of processing activities that regulators can independently audit.
Compliance is therefore not an external feature bolted onto the application, but an end-to-end property that must extend across the entire workflow. 
Meeting this goal demands three key requirements: \emph{(i)}~confidential and integrity‑protected personal data processing and storage; \emph{(ii)}~policy-aware KV operations with verifiable audit logging; and \emph{(iii)}~preservation of the performance and simplicity of KVS.

\section{Data Security \& \gdpr Compliance Challenges}
\label{sec:motivation}

KVSs are architected for simplicity and performance, fundamentally complicating the verifiable enforcement of \gdpr requirements~\cite{gdprbench}. Meeting these requirements ($\S$~\ref{sec:background}) in untrusted cloud environments introduces data management challenges across three dimensions: \emph{security}, \emph{policy specification and enforcement}, and \emph{performance}.

\subsection{Security in Untrusted Cloud Environments}
\label{subsec:sec_challenges}

KVS-based applications in untrusted clouds face \gdpr risks because privileged operators can access or tamper with in-memory KV pairs, stored data, operation logs, and audit records. Conventional protections such as authentication, access-control lists, or TLS encryption---while effective for data in transit or at rest---implicitly assume a trusted operating system and hypervisor, assumptions that no longer hold in public cloud environments.
Therefore, achieving \gdpr compliance requires end-to-end protection covering both volatile state (e.g., in-memory data, policy decisions, compliance metadata) and persistent state (e.g., KV pairs and audit logs).

~%
CVMs establish a strong security foundation by offering three core guarantees: confidentiality of volatile state, ensuring data and metadata privacy; integrity protection, which prevents unauthorized modification of data and policies during execution; and remote attestation, allowing for external verification of the deployment.
However, placing a KVS in a CVM alone is insufficient for \gdpr compliance because the latter depends on the integrity of policy enforcement and metadata management throughout the workflow.  Simply encrypting values does not ensure that deletion requests are honored, retention and purpose policies are correctly enforced, or audit logs remain authentic once data leaves the trusted boundary.

{\em The core security challenge} is thus to extend CVM guarantees beyond volatile state to persistent storage and logs, so that policy decisions and compliance records can be verified end-to-end. These guarantees must remain general and compatible across heterogeneous KVS without altering their data models. This requires orchestrating KV operations and dataflows so that data, metadata, and policy enforcement execute in a unified, verifiable environment. %

\subsection{Policy Specification and Enforcement}
\label{subsec:compl_challenges}

\gdpr defines a broad legal framework for data protection, but not all of its articles are equally relevant to the KVS layer~\cite{gdprbench}.
We focus on requirements technically enforceable within the KVS, such as ``purpose limitation'' and ``storage limitation'' (Article 5), ``right of access'' (Article 15), ``right to be forgotten'' (Article 17), ``right to object'' (Article 21), and ``records of processing activities'' (Article 30). 
Articles handled at higher application layers---such as breach notification (Articles 33-34) and data portability (Article 20)---are data controller responsibilities that extend beyond the KVS scope.

\gdpr obligations relevant to KVS introduce {\em policy specification challenges}, as \gdpr's legal language cannot be easily mapped to technical constraints. %
Unlike relational databases with structured schemas and rich metadata support, KVS offers limited expressiveness for defining complex policies. %
For example, determining if user behavior tracking for ``improving user experience'' resembles the same purpose (Article 5) as ``product optimization'' requires legal interpretation that KVS systems cannot handle autonomously.  
This gap between legal requirements and technical implementation creates uncertainty on whether systems truly satisfy the \gdpr rules, exposing organizations to compliance risks despite their efforts. 

Beyond specification, {\em policy enforcement introduces its own challenges}. 
\gdpr mandates associating user data with rich compliance metadata that govern how it can be accessed or modified.
In practice, ensuring correct and efficient enforcement across KVS operations is non-trivial.
As KVS lack structured schemas or relational integrity, maintaining key-policy associations through updates or cache operations can introduce inconsistencies. 
Moreover, the granularity of enforcement, whether at the level of individual keys or namespaces, 
can cause \emph{metadata proliferation}, demanding careful trade-offs between scalability, consistency, and resource efficiency. 
These factors collectively make it challenging to realize correct and efficient system-level enforcement of regulatory policies with modern KVS.

\subsection{Preserving Performance}
\label{subsec:perf_challenges}

\gdpr mandates rights (e.g., data access, retention) that force KVS to locate and manage records via \gdpr metadata, not simple key lookups. These metadata-driven queries (e.g., retrieving all data of a user, implementing the ``right to be forgotten'') are inefficient for existing KVS, causing severe latency and I/O overheads. For example, naïve implementations in Redis show latencies of several seconds even on modest datasets~\cite{gdprbench}.

Beyond query processing, compliance mechanisms themselves introduce performance penalties: \gdpr's audit logging demands creating metadata-rich entries for every operation on sensitive data, transforming lightweight requests into heavier I/O workloads with multiple physical writes. %
Structured \gdpr audit logs also require more complex serialization and parsing than simple append-only logs, adding CPU overhead that compounds the I/O bottleneck and erodes the latency advantages of KVS architectures.

Addressing these {\em performance challenges} requires rethinking how \gdpr mechanisms integrate with the KVS data path. 
A practical approach involves representing compliance metadata compactly, indexing it to avoid full KVS scans, and recording audit trails efficiently without disrupting normal KV operations. Equally important is minimizing interference with the critical data path: reads and writes must remain lightweight, while policy enforcement should be optimized through efficient validation pipelines.

\noindent\fbox{\parbox{\columnwidth}{
\myparagraph{Problem statement} To address these challenges, we target the following question: \emph{How can we design a system that enables secure \gdpr compliance for KVS on untrusted clouds while maintaining high-performance query processing, providing verifiable policy enforcement, and preserving compatibility with existing deployments?} 
}}

\section{Overview}

As a solution, we present \projectname{}, a trusted middleware system that acts as a \gdpr monitor to transparently ensure \gdpr compliance for cloud KVS through strict policy enforcement, efficient \gdpr metadata management, and verifiable audit logs.

\subsection{System Architecture}
\label{sec:arch}

\begin{figure}[t]
\centering
\includegraphics[width=0.8\columnwidth]{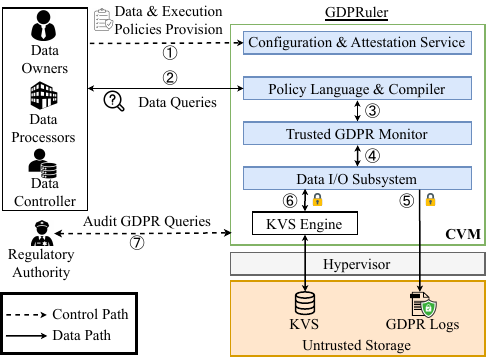}
\vspace{-1mm}
\caption{\projectname{} architecture and workflow.}
\label{fig:gdpruler-overview}
\vspace{-2mm}
\end{figure}

\autoref{fig:gdpruler-overview} depicts \projectname{}'s architecture, highlighting its trusted boundary (green box) and key components. 
Interposed between applications and unmodified KVS engines, it runs within a CVM that establishes an isolated, attested environment mediating all KV operations. 
Within this boundary, it enforces \gdpr policies, validates query intents, and maintains tamper-evident compliance logs, while preserving compatibility with existing KVS.
\projectname{}'s design is guided by three principles tailored to KVS traits:

\begin{itemize}[noitemsep,leftmargin=*]
    \item \textbf{Policy-driven query control.} 
    Since KVS lack schemas or query planners for access controls,
    \projectname{} introduces a \textit{policy specification language} and \textit{companion API} for  data owners and processors to express policies alongside KV queries. The owners define how personal data must be managed (e.g., consent), %
    while processors declare their intended purposes. These policies are compiled into compact metadata structures attached to queries, transforming opaque KVS requests into policy-aware operations.%

    \item \textbf{Trusted enforcement and verifiable auditing.} 
    \projectname{} enforces compliance within a \textit{trusted monitor} running in a CVM. All compliance decisions remain confined to this protected domain. A \textit{KVS-tailored tamper-evident log} records policy-relevant actions, while \textit{remote-attestation protocols} extend trust to external \gdpr entities
    allowing them to verify that \gdpr enforcement and auditing originate from an authentic, untampered deployment.

    \item \textbf{Backend-agnostic deployment and extensibility.} 
    \projectname{} operates as a drop-in middleware that can sit atop any KVS. A \textit{unified interface layer} abstracts backend-specific APIs, enabling integration with KVS (e.g., Redis) without re-implementing compliance logic. The design supports future backends through lightweight adapters and performance-oriented optimizations, such as efficient metadata encodings, and metadata indexing (see \autoref{sec:design}).
\end{itemize}

Guided by these principles, \projectname{} comprises five modular components. The \textit{Configuration and Attestation Service} anchors system trust by managing entity registration and establishing secure channels via mutual remote attestation, ensuring that only legitimate parties interact with the system and register their default data and execution policies used for subsequent enforcement. 

The \textit{Policy Compiler} translates these policies to executable rule sets optimized for KVS workloads, intercepting KV operations and transforming them to \gdpr-aware queries carrying compliance metadata. It applies compact encodings and merges default policies with per-query overrides for fine-grained runtime enforcement. 

The \textit{Trusted GDPR Monitor} enforces the compiled rules within the CVM, validating access, modification, and retention constraints while managing the tamper-evident log. 
For data operations, it performs access control and filters results to ensure that the requesting entity's declared policy aligns with owners' consents, sharing permissions, and retention settings. 
It maintains auxiliary data structures optimizing \gdpr queries and runs a background expiration scanner. %

The \textit{Data I/O Subsystem} bridges the monitor and the backend engine, encrypting payloads, authenticating operations, and ensuring the confidentiality and integrity of both data and logs. 
It generates tamper-evident records for all policy-relevant actions and exposes an authenticated audit trail accessible to regulators. 

Finally, the \textit{KVS Engine} executes standard key--value operations over encrypted values without GDPR awareness, ensuring compatibility with unmodified backend software. 
Both encrypted data files and audit logs reside in untrusted storage outside the CVM.

\subsection{System Workflow}
\label{subsec:workflow}

\myparagraph{1. Setup phase}
Before launch, the data controller registers public keys of all legitimate entities---data owners, processors, controllers, and regulatory authorities---with the \textit{Configuration and Attestation Service}, establishing the system's trust anchors. 
Entities verify \projectname{}'s integrity through remote attestation, while the service validates their cryptographic identities against the registered keys. 
After successful mutual attestation, secure channels are established, and session keys are issued for subsequent interactions. 
Data owners and processors then upload default data and execution 
policies to the \textit{Trusted \gdpr Monitor}, securing them inside the CVM (\Circled{1}).

\myparagraph{2. Query execution}
When an entity issues a query (\Circled{2}), it may include \gdpr predicates that act as complementary constraints. 
The \textit{Policy Compiler} parses these predicates and merges them with the entity’s default policy, generates an enforcement rule, and attaches the corresponding metadata to the query. 
The augmented operation is passed to the \textit{Trusted \gdpr Monitor} (\Circled{3}), which validates it against the compiled rules, and forwards it to the \textit{Data I/O Subsystem} (\Circled{4}).

\myparagraph{3. Data processing and logging}
The \textit{Data I/O Subsystem} creates a tamper-evident log entry for each operation involving personal data or in case of a policy violation (\Circled{5}). 
Then, it encrypts the payload (data, metadata), maps the operation to the appropriate API (e.g., Redis \emph{SET/GET}, RocksDB \emph{Put/Get}), and submits it to the \textit{KVS} (\Circled{6}) for execution.
Subsequently, it returns the result to the monitor, which performs final validation before delivering it to the requesting entity.

\myparagraph{4. Auditing}
For regulatory inspection, authorized authorities can request access to \gdpr logs (\Circled{7}), enabling verification of compliance and detection of possible violations. The \textit{Trusted \gdpr Monitor} authenticates the requester based on its registered key.

\if
\myparagraph{\#1: Complete and verifiable \gdpr compliance}
\projectname{} provides a comprehensive policy language composed by a well-defined set of predicates that is sufficient to express the articles of \gdpr. 
Additionally, it contains a dedicated logging layer to produce tamper-evident logs for the users' private data that can be audited by the regulatory authorities and prove the \gdpr compliance or report potential data breaches.
Lastly, \projectname{} leverages TEEs (i.e., CVMs) that, apart from ensuring the security guarantees (i.e., integrity, confidentiality) of the data in use, enables external parties to gain trust in the deployed system and attest its state prior to proceeding with secret sharing.  

\myparagraph{\#2: Performance and storage overheads} \projectname{} strives to maintain the performance overheads incurred by the additional \gdpr metadata handling and the extensive logging of critical query operations on private data at a reasonable level. On top of that, to cope with the metadata explosion, \projectname{} carefully encodes the \gdpr information and performs optimizations on the logging layer (e.g., compression) to reduce the required storage space.

\myparagraph{\#3: Transparency and ease of use}
\projectname{} provides a generic and transparent monitor suitable for existing KV database backends. It is responsible to embed and handle the \gdpr metadata, perform the access control and the logging mandated by the \gdpr rules without requiring internal code modifications in the data management engine or specialized knowledge from the developer.

1. How to create a generic and transparent monitor to existing database backends while embedding the \gdpr metadata.. No uniform and secure way of enforcing the GDPR policies without redesigning the backend database engine
2. How to provide a complete and externally verifiable \gdpr compliance (including logging etc.) with a simple set of predicates and frontend API that can be easily used
3. Ensure end-to-end data security (confidential VMs + attestation)
4. Minimize performance and space overheads to make the system usable (see optimization section in implementation...) especially for NoSQL DBs (see GDPRBench 6.2)
\fi

\vspace{-4mm}

\begin{center}
\begin{table}
\footnotesize
\begin{tabular}{c||l} 
 \hline
 \textbf{Query API} & \textbf{Description}\\
 \hline
 \emph{get(k)}  & Retrieves the \emph{value} associated with \emph{k}.\\
 \hline
 \emph{put(k,v)}  & Inserts a KV pair and its metadata.\\
 \hline
 \emph{delete(k)}  & Deletes the specified KV pair.\\
 \hline
 \emph{getm(key\_prefix,}  & Retrieves \emph{values} or metadata. Results are\\
 \emph{~~~~~~~~~~~~~~~~~~~~~~~~``(meta)data'')} & filtered based on the  query policy.\\
 \hline
 \emph{putm(key\_prefix,}  & Updates \gdpr metadata for KV pairs\\
 \emph{~~~~~~~~~~~~~~~~~~~~~~~~new\_metadata)} & that comply with the query policy. \\
 \hline
 \multirow{1}{*}{\emph{deletem(key\_prefix)}}  & Deletes KV pairs that comply with the query policy.\\
 \hline
 \emph{getLogs(optional\_key)}  & Retrieves all \gdpr logs (or those of a specified key).\\
 \hline
\end{tabular}
\caption{\projectname{}'s Query API.}
\label{tab:api}
\vspace{-4mm}
\end{table}
\end{center}   

\subsection{Programming Model}
\label{sec:prog_model}

\projectname{} exposes a standard \emph{put/get/delete} API, summarized in \autoref{tab:api}. %
To support diverse backends that may use distinct bindings, \projectname{} provides adapters that map these operations to its unified API.
To support metadata management and auditing, required for \gdpr compliance (\autoref{subsec:compl_challenges}), \projectname{} provides \emph{compliance-aware operations} (\emph{getm}/\emph{putm}/\emph{deletem}) and a logging primitive (\emph{getLogs}), enabling the management of \gdpr-related metadata and audit logs.%

Before issuing queries, each GDPR entity specifies a default data handling policy, as shown in the listing below. %
\begin{lstlisting}[style=customc]
// Data owner (Alice) default data policy:
{"sessionKey":alice_key, "default_policy": {
  "purpose":[recommendations,orders], "share":[recommender_key],
  "objection":[marketing,analytics], "expTime":[90d],
  "origin":[shop.com/account_creation], "monitor":[true]}}
// Data processor (Recommendation Service) default execution policy.
{"sessionKey":recommender_key,
 "default_policy": {"purpose":[recommendations]}}
\end{lstlisting}

Entities express queries with \projectname{}'s policy specification language (\autoref{subsec:policy_language}) using predicates shown in \autoref{tab:lang-predicates}.
If no per-query predicates are specified, the system automatically applies the entity’s default policy. %
The following listing presents query examples.
\begin{lstlisting}[style=customc,mathescape]
// Data owner: store data using the default policy.
query(put("alice:preferences","data")) $\wedge$ sessionKey(alice_key)
// Data processor: request data for the "recommendations" purpose.
query(get("alice:preferences")) $\wedge$ sessionKey(recommender_key)
   $\wedge$ objPurIs(recommendations)
\end{lstlisting}

\begin{figure*}[t]
\centering
\includegraphics[width=0.75\textwidth]{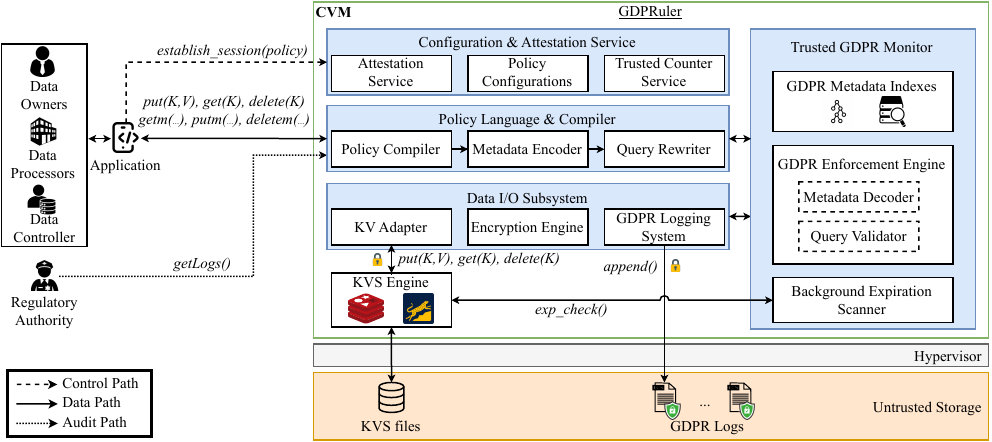}
\vspace{-1mm}
\caption{Detailed view of \projectname{} system. \projectname{} consists of \emph{(i)}~a policy compiler for \projectname{}'s declarative policy language, \emph{(ii)}~the Trusted \gdpr Monitor, \emph{(iii)}~\projectname{}'s Data I/O Subsystem, and \emph{(iv)}~Configuration and Attestation Services.}
\vspace{-5mm}
\label{fig:gdpruler-design}
\end{figure*}

\subsection{Deployment and Threat Model}
\label{subsec:deployment-threat}

\projectname{} assumes a platform with CVM support to host its trusted components, such as AMD SEV-SNP~\cite{sev-snp}, Intel TDX~\cite{tdx}, and ARM CCA~\cite{cca}. %
External applications operated by data owners, processors, controllers, and regulators interact with \projectname{} over mutually attested, encrypted channels established during system setup.

Applications outside \projectname{} are assumed to run within trusted domains (e.g., CVM, on-premises environment). 
Securing application logic is beyond our scope. 
The system uses persistent storage for KVS data and compliance logs. %
We assume this storage provides durability but not rollback protection. 
Deployments may instantiate a single CVM instance or multiple ones %
depending on scalability needs.

Under this deployment setting, the trusted computing base (TCB) of \projectname{} comprises the CVM hardware, its firmware, and all software components running inside the CVM. 
All other system software and infrastructure---including the hypervisor and external storage---is untrusted. 
\projectname{} inherits the standard adversary model of modern CVMs~\cite{sev-snp,sev-snp-abi,tdx2,tdx3}, defending against powerful attackers~\cite{dolev_yao} that can compromise the system software stack or perform physical attacks such as memory probing, while assuming the processor package and on-chip state remain protected.

We do not consider side-channel attacks~\cite{alam_cc-side-channel_2023,intel_secure_coding_guideline,wichelmann_cipherfix_2023,sev-cipher-sidechannel-mitigation,deng_cipherh_2023,firmware-sidechannel-mitigation,wang_pwrleak_2023}, denial-of-service, or timing manipulation (e.g., clock/frequency tampering~\cite{securetsc_issues}). 
For \gdpr logs, we consider adversaries capable of modifying the data or launching rollback and forking attacks~\cite{rollback_parno,rote,forking_lcm}.

\section{Design}
\label{sec:design}

\begin{table}[t]
\footnotesize
\setlength\tabcolsep{0pt}
\centering
\begin{tabular}{p{3.5cm} p{5.0cm}}
\hline
\multicolumn{2}{c}{\bfseries Relational predicates (for \projectname{} policy enforcement)} \\
\hline
\textit{eq}(x,y), \textit{le}(x,y), \textit{lt}(x,y) & x $=$ y, x $\leq$ y,  x $<$ y \\
\textit{ge}(x,y), \textit{gt}(x,y) & x $\geq$ y, x $>$ y\\
\hline
\multicolumn{2}{c}{\bfseries Session predicates} \\
\hline
\textit{sessionKey}($K$) & Sets and/or represents the session key K \\
\hline
\multicolumn{2}{c}{\bfseries GDPR metadata predicates} \\
\hline
\textit{objOwn}($x$), \textit{objOwnIs}($x$) & Represents the owner of an object \\
\textit{objExp}($x$), \textit{objExpIs}($x$) & Represents the expiry time of an object \\
\textit{objPur}($x$), \textit{objPurIs}($x$) & Represents the purposes for an object or an entity\\
\textit{objOrig}($x$), \textit{objOrigIs}($x$) & Represents the origin of an object \\
\textit{objShare}($x$), \textit{objShareIs}($x$) & Represents the sharing information of an object \\
\textit{objObj}($x$), \textit{objObjIs}($x$) & Represents the objections imposed on an object \\
\hline
\multicolumn{2}{c}{\bfseries GDPR data predicates} \\
\hline
\textit{monitor}($true/false$) & Enables/disables monitoring \\
\textit{encryption}($true/false$) & Enables/disables encryption \\
\hline
\multicolumn{2}{c}{\bfseries Query predicates} \\
\hline
\textit{query}($put/get/delete$) & Represents a standard query operation \\
\textit{query}($putm/getm/deletem$) & Represents a \gdpr query operation \\
\textit{query}($getLogs$) & Represents a \gdpr getlog query operation \\
\hline
\end{tabular}
\caption{\projectname{} policy language predicates.}
\vspace{-2mm}
\label{tab:lang-predicates}
\end{table}

\subsection{Policy Language and Compiler}
\label{subsec:policy_language}
The Policy Language and Compiler transforms data policies into \gdpr-aware operations.
We first describe how policy is specified with our novel declarative language, along with associated metadata, and then detail its compilation process for enforcement.

\myparagraph{\gdpr metadata}
In \projectname{}, each KV pair (object) has associated metadata that keeps the policies necessary for \gdpr compliance.
This metadata includes its \emph{owner} (\emph{session key}), \emph{origin} (data source), \emph{purpose} (allowed uses), \emph{objections} (prohibited uses), \emph{sharing permissions} (which processors may access the data),  \emph{expiration date}, an \emph{encryption} flag, and a \emph{monitor} flag (logging on/off).
This metadata is created when a value is inserted and is either provided along with the insertion query or initialized with default policies.
Internally, the metadata is prepended to the associated value of a KV pair and stored in the KVS.
Besides \gdpr metadata associated with KV pairs, \projectname{} preserves metadata for data processors (i.e., session key, data processing purposes), initially provided during the setup phase.

\myparagraph{Declarative policy language}
Drawing inspiration from declarative language for data policy specification (e.g., \cite{pesos,ironsafe,guardat}), \projectname{} defines a policy language that enables data owners and processors to specify their data-handling policies along with KV queries in a formal predicate-based syntax (\autoref{tab:lang-predicates}).
These policies use session predicates to specify their session, \gdpr metadata and \gdpr data predicates to specify \gdpr policies, and the query predicate to specify the actual query, with multiple predicates joined with `$\wedge$'.
For instance, a typical query is formed like the following:
\begin{lstlisting}[style=customc,mathescape]
query(op) $\wedge$ sessionKey(K) $\wedge$ objExp(x) $\wedge$ objPur(y) $\wedge$ ...
\end{lstlisting}

\projectname{} policies are categorized into two parts: \emph{data access} and \emph{execution} policies.
Data access policies override the data owner's default policy.
Specifically, preference type predicates (predicates of the form \emph{objXXX()}, e.g., \emph{objPur(x)}) explicitly specify the policy associated with the KV pair.
For example, Alice (data owner) uses
\begin{lstlisting}[style=customc,mathescape]
query(put("alice:purchase","data")) $\wedge$ sessionKey(alice_key) $\wedge$
    objPur(orders) $\wedge$ objObj(recommendations)
\end{lstlisting}
to insert ``data'' with a key ``alice:purchase'' using ``alice\_key'' as a session key with the purpose of ``orders'' and objection to ``recommendations''.
Additionally, filter type predicates (predicates of the form {\emph{objXXXIs()}}) are used to create a filter for the associated metadata (referred to as the \emph{metadata filter}) when performing \gdpr queries (\emph{getm/putm/deletem}).
For example, the following operations get values whose key prefix is ``alice:'', and their purpose is ``orders''.
\begin{lstlisting}[style=customc,mathescape]
query(getm("alice:","data"))$\wedge$sessionKey(alice_key)$\wedge$objPurIs(orders)
\end{lstlisting}

On the other hand, execution policies control the intent of data processors.
With this policy, for instance, the data recommendation service (data processor) may send the following query intended to get a value of ``alice: purchase'' for recommendation purposes.
\begin{lstlisting}[style=customc,mathescape]
query(get("alice:purchase")) $\wedge$ sessionKey(recommender_key) $\wedge$
    objPurIs(recommendations)
\end{lstlisting}

\myparagraph{\projectname{} policy enforcement rules}
The Trusted \gdpr Monitor uses an extended version of our policy language to describe its policy enforcement rules.
They are expressed in the form \textit{perm~:-~condition}.
The \textit{condition} is composed of relational predicates (e.g., \emph{eq(x,y)}) and predicates used by data access and execution policies.
Multiple relational predicates are joined with `$\wedge$' (logical and) or `$\vee$' (logical or) operators. 
The filter type predicates (predicates of the form \emph{objXXXIs()}) are extended to return metadata values associated with a session key or object (e.g., \emph{objPurIs(K)} returns the purpose of the entity K).
The language also defines `\&' (intersection) and `||' (union) operations for metadata values.
Using this policy language, \projectname{} clearly specifies the \gdpr enforcement rules in a declarative manner.
For instance, the following enforces the purpose limitation (Article 5):
\begin{lstlisting}[style=customc,mathescape=true]
// objPurIs(K): the purposes of the entity with session key K
// objPurIs(k): the allowed purposes of the data object with key k
$\textbf{read}$ :- eq(objPurIs(K) & objPurIs(k), objPurIs(K))
\end{lstlisting}
We detail how the predicates form \gdpr enforcement rules in \autoref{section:gdpr_use_cases}.

\myparagraph{Policy compiler}
The policy compiler transforms high-level \gdpr policy expressions into executable enforcement directives. %
For each query operation, it parses the policy predicates from the request, tokenizes these expressions, and performs semantic analysis to extract predicate attributes and values.
It then combines those attributes with default policies to fill the missing values.
For \gdpr queries, it generates \emph{metadata filters} for later filtering operations.

\myparagraph{Metadata encoder}
For \emph{put/putm} operations, metadata is stored along with the associated values.
\projectname{} utilizes bitmaps for values that represent categories (e.g., purposes), enabling quick bitwise operations for policy validation.
The mapping of bits is established during runtime, and all bitmap structures can be resized to accommodate new entries. This ensures the system remains adaptable to changing compliance requirements while minimizing metadata and maintaining efficient processing.
The expiration times have 8-byte precision to ensure precise time-based policy enforcement.

\myparagraph{Query rewriter}
Given the policies from the policy compiler, the query rewriter
transforms standard KV queries into \gdpr-aware operations.
For write operations (e.g., \textit{put}), it prepends the encoded \gdpr metadata to the user data. %
For read operations (e.g., \textit{get}), it propagates the query along with the associated metadata filters to the \gdpr Enforcement Engine for subsequent validation.

\subsection{Trusted \gdpr Monitor}
\label{subsec:des:secure_runtime}

The Trusted \gdpr Monitor serves as the central compliance verification component of \projectname{}.
It processes \gdpr-aware operations and ensures all access complies with specified policies. %

\myparagraph{\gdpr enforcement engine}
The \gdpr enforcement engine receives operations with \gdpr metadata from the query rewriter.
For \emph{get/getm} operations, it also retrieves (encoded) \gdpr metadata of the associated object, and the \emph{Metadata Decoder} decodes it.
Subsequently, its \emph{query validator} validates the policy enforcement rules. After successful validation, it forwards the operations to the Data I/O Subsystem to materialize the KV operation and logging.

The query validator %
first confirms that the data is not expired and validates that the request initiator either is the data owner (for updates) or has explicitly been granted access via the sharing property (for reads). Then, it checks that declared purposes fall within the subset of allowed purposes, and finally verifies that no query purposes are part of the KV pair's objections.

\begin{figure}[t]
\centering
\includegraphics[width=0.8\columnwidth]{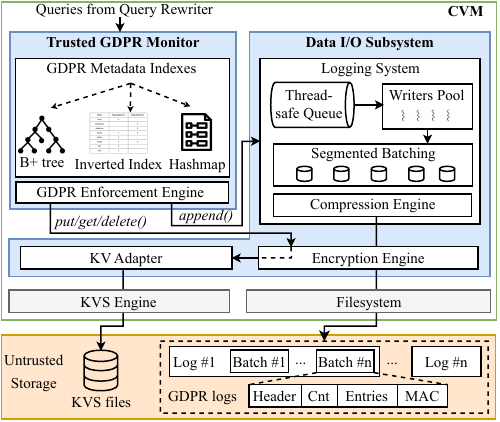}
\vspace{-1mm}
\caption{Metadata indexing and data I/O subsystems.}
\vspace{-1mm}
\label{fig:gdpruler-metadata-index}
\end{figure}

\myparagraph{\gdpr metadata indexes}
To optimize policy validation,
\projectname{} maintains specialized in-memory data structures that index keys based on their \gdpr metadata, as shown in ~\autoref{fig:gdpruler-metadata-index}, avoiding full KVS scans for metadata operations.
\projectname{} provides different indexing options (e.g., hashmap, B+ tree, inverted index). %
Data controllers can enable specific data structures based on anticipated \gdpr query patterns and metadata types. 
For instance, hashmap indexes provide O(1) lookups for exact metadata matches, while B+ trees support efficient range queries, useful for temporal fields. %

\myparagraph{Background expiration scanner}
\projectname{} implements a background expiration scanner to enforce \gdpr's data minimization and storage limitation principles (Article 5). 
The scanner operates asynchronously at configured intervals. %
It scans metadata and queues delete requests on expired data to the Data I/O Subsystem.
While the KVS backend handles actual deletion, which may implement techniques such as tombstoning~\cite{rocksdb-paper}, \projectname{} ensures that expired data becomes unreachable through its API via its \gdpr Enforcement Engine.
This approach represents a reasonable trade-off for \gdpr compliance, while respecting KVS-specific deletion mechanisms.

\subsection{Data I/O Subsystem}
\label{subsec:des:data_layer}

\projectname{}'s \gdpr Data I/O Subsystem handles all data encryption, KV backend communication, and tamper-evident audit logging.

\myparagraph{KV adapter} The KV adapter bridges \projectname{}'s components with the \emph{unmodified} KVS engine. 
It implements a template-based approach where core operations (e.g., \emph{put}, \emph{get}, \emph{delete}) are mapped to the specific API of different KV backends. This design enables the seamless integration of diverse KV stores by utilizing generic data types (e.g., byte streams) that are compatible with most implementations.

\myparagraph{Encryption engine} 
To ensure confidentiality (Article 32), even if a KVS does not inherently support encryption, \projectname{} offers a built-in encryption engine. 
During connection establishment, data owners and regulatory authorities perform authenticated key exchange with \projectname{} using their session keys to derive encryption keys.
The encryption engine encrypts user data (values) and log entries and produces a Message Authentication Code (MAC) as part of each object's metadata to validate integrity upon decryption.

\myparagraph{\gdpr logging system}
To address the \gdpr auditing requirements (Article 30/33), \projectname{} incorporates a secure, tamper-evident logging system (\autoref{fig:gdpruler-metadata-index}) that ensures the confidentiality, integrity, and freshness properties of \gdpr logs. 
Log entries are batched and compressed to optimize write throughput and reduce write amplification.
Specifically, they are enqueued in a thread-safe queue, and processed by dedicated writer threads. %
Processed entries are grouped as a batch based on a hash of the affected key.
The batch contains a metadata header, a monotonically incremented software-based trusted counter protected within the CVM for freshness verification, a series of \gdpr entries encrypted along with the counter, and a MAC for integrity checks. 
The batched log is written to append-only files, and log files are automatically rotated based on configurable size thresholds. %

For log access requests from regulatory authorities, \projectname{} decrypts the identified files per-batch, verifies integrity (via MAC), and confirms monotonically increasing counter values within the expected range, thereby preventing replay attacks.

\subsection{Configuration and Attestation Service}
\label{subsec:des:attestation_and_configuration}

The Configuration and Attestation Service establishes trust between external entities and the \projectname{} instance, manages policy configurations, and maintains session security.

\myparagraph{Attestation service}
\projectname{}'s attestation service enables external entities to ensure the integrity of a \projectname{} and KV engine before establishing secure communication channels. They verify the platform configuration, guest VM measurements, and \projectname{} binaries through attestation reports generated by secure hardware (e.g., AMD Secure Processor~\cite{sev-snp-abi}). %
Following successful verification, \projectname{} performs application validation against entity keys, pre-registered by the data controller, ensuring that only pre-authorized, legitimate applications can establish connections with it. %

\myparagraph{Policy configurations}
The policy configuration component manages default \gdpr policies, which are provided by each entity after successful mutual attestation.
The default policies are associated with the entity's session key and stored within the CVM memory. 
\if
The policy configuration component manages the storage and retrieval of \gdpr policies.
Upon successful mutual attestation, each entity provides \projectname{} with their desired default data or execution policies. The policy confirmation component associates them with the entity's session key and stores them within the CVM. 
Importantly, these default policies can be overwritten on a per-query basis, where specified metadata (e.g., updated purpose of use) and conditional properties (e.g., data owner) take precedence over the default policy, while unspecified values default to the original policy settings. \masa{this is already explained in 5.1}
\fi

\myparagraph{Trusted counter service}
The trusted counter service provides monotonically increasing counters %
to ensure the freshness of the audit logs.
It maintains a software-based counter for each log file that increments with each appended batch.
During log retrieval, \projectname{} verifies that the log's counter value matches the expected sequence. %
Any deviation from the expected values indicates tampering or rollback attacks.
The counters are periodically checkpointed to encrypted storage to survive system restarts. %

\begin{algorithm}[t]
\caption{Unified GDPR-Aware Operation Framework}
\label{alg:unified-gdpr}
\myfontsizealgorithms
\SetAlgoLined
\SetAlgoSkip{}
\SetKwInOut{KwIn}{Input}
\SetKwInOut{KwOut}{Output}
\KwIn{~$operation$,$key/key\_prefix$,$value$,$session\_key$,$query\_policy$}
\KwOut{~Compliance-validated result or denial of operation}

\underline{\textbf{Stage 1: Policy Compilation}} \\
$default\_policy \leftarrow$ \texttt{get\_session\_policy}($session\_key$); \\
$policy \leftarrow$ \texttt{compile\_query\_policy}($default\_policy$, $query\_policy$); \\

\underline{\textbf{Stage 2: Data Retrieval and Metadata Extraction}} \\
$encrypted\_value(s) \leftarrow$ \texttt{kv\_store.get}($key$ or $key\_prefix*$); \\
$value(s)\leftarrow$ \texttt{decrypt}($encrypted\_value(s)$); \\
$gdpr\_metadata \leftarrow$ \texttt{extract\_metadata}($value(s)$); \\

\underline{\textbf{Stage 3: Access Validation and Compliance Checking}} \\
\ForEach{$entry \in$ $gdpr\_metadata$}{
    $compliance \leftarrow$ \texttt{validate}($entry$, $policy$, $session\_key$); \\
    \If{$\neg compliance$}{
        \texttt{monitor\_query}($key$, $operation$, $compliance$); \\
        Skip or deny this $entry$; \\
    }
}

\underline{\textbf{Stage 4: Operation Execution and Audit Logging For Each Value}} \\
\uIf{$operation \in \{put, putm\}$}{
    $enhanced\_value \leftarrow$ \texttt{rewriter.create\_new\_value}( $policy$, $value$); \\
    \If{$gdpr\_metadata.encrypt$}{
        $enhanced\_value$ = \texttt{encrypt}($enhanced\_value$);
    }
    $result \leftarrow$ \texttt{kv\_store.put}($key$, $enhanced\_value$); \\
}
\uElseIf{$operation \in \{get, getm\}$}{
    $result \leftarrow$ \texttt{filter\_and\_aggregate}($compliant\_entries$); \\
}
\uElseIf{$operation \in \{delete, deletem\}$}{
    $result \leftarrow$ \texttt{kv\_store.delete}($compliant\_keys$); \\
}
\If{$gdpr\_metadata.monitor$ $\vee$ $violations detected$}{
    \texttt{monitor\_query}($key$, $operation$, $result$, $compliance$); \\
}
\Return $result$;
\end{algorithm}

\subsection{Algorithms}

\autoref{alg:unified-gdpr} illustrates how \projectname{} converts standard KV operations to \gdpr-compliant operations through a four-stage pipeline:
\emph{(i)}~policy compilation, and merging default and query-specific predicates, \emph{(ii)}~metadata retrieval from the object, \emph{(iii)}~policy validation for compliance checks, and \emph{(iv)}~operation execution with audit logging.
\ifthenelse{\boolean{appendix}}
{Detailed algorithms for all operations are covered in Appendix~\ref{sec:appendix:algos}.
}
{Detailed algorithms for all operations are covered in Appendix A.
} 

\myparagraph{Standard KV queries}
For {\em get} operations, \projectname{} first compiles policies from the requesting entity and any operation-specific overrides (Stage 1), retrieves and decrypts the KV pair and extracts embedded \gdpr metadata (Stage 2), validates access permissions (e.g., ownership, purposes, objections) (Stage 3), and conditionally logs the access if required by the data's privacy settings (Stage 4). Failed access attempts are always logged for audit purposes. %

{\em put} operations execute Stages 1 and 2 of the pipeline to check if a key exists and fetch its current metadata. To reduce KVS stress, \projectname{} preserves an in-memory cache associating a subset of keys with their metadata. For existing keys, Stage 3 validates modification permissions based on their current metadata. In Stage 4, the query rewriter encodes metadata directly within the value, creating a payload that embeds \gdpr properties. 
The operation is then submitted to the KVS and gets conditionally logged.

{\em delete} operations follow a similar pipeline to \emph{get/put}. %
\projectname{} ensures that deletion rights are properly enforced while maintaining comprehensive audit trails.%

\myparagraph{\gdpr queries}
For {\em getm} operations, \projectname{} performs bulk retrieval using key prefix matching (Stage 2), then applies a two-stage filtering: (1) filters entries matching specified criteria, and (2) performs individual \gdpr compliance validation. %
In Stage 4, every data access is logged, if required, creating detailed audit trails.%

The {\em putm} operation executes Stages 1 and 2 to compile policies and perform bulk retrieval. In Stage 3, the system applies \gdpr metadata filters to identify entries matching the specified criteria, then validates modification permissions for each entry. %
Stage 4 leverages the query rewriter to create enhanced values.

The {\em deletem} operation follows the unified framework adapted for bulk deletion. After policy compilation (Stage 1) and bulk retrieval (Stage 2), Stage 3 validates that the requesting entity has deletion rights for each matching pair strictly based on ownership. 
In Stage 4, \projectname{} performs bulk deletion of compliant KV pairs and logs each deletion with comprehensive audit information.

\myparagraph{Regulatory accesses}
The {\em getLogs} operation requires explicit regulator authentication and follows a specialized retrieval path. The operation validates regulatory authority credentials through their registered keys, temporarily pauses ongoing logging to ensure consistency in the retrieved audit trail, retrieves log files either for specific keys (using hash-based partitioning) or the entirety of the log files, and internally performs integrity and freshness verification and reports any tampering detections to the authorities.

\subsection{ \gdpr Policy Predicates}
\label{section:gdpr_use_cases}
\begin{table}
\footnotesize
\begin{tabular}{l|| c } 
 \hline
 \textbf{\gdpr Articles} & \textbf{Policy enforcement rule}\\
 \hline
 \makecell[l]{\emph{\#5 Purpose limitation}}  & \makecell[l]{\textbf{read}{:}{-} \text{eq}(\text{objPurIs}($K$) \& \text{objPurIs}($k$), \text{objPurIs}($K$))}\\
 \hline
 \makecell[l]{\emph{\#5 Storage limitation}}  & \makecell[l]{\textbf{read}{:}{-} \text{le}(\text{time}, \text{objExpIs}($k$))}\\
 \hline
 \makecell[l]{\emph{\#13,\#14 Information to }\\ \emph{be provided}}  & \makecell[l]{
 \textbf{read\_{metadata}}{:}{-}
  eq(\text{objOwnIs}($k$),   \text{objOwnIs}($K$))  \\ 
 } \\
 \hline
 \makecell[l]{\emph{\#15 Right of access by users}}  &
  \makecell[l]{\textbf{read(\_{metadata})}{:}{-} eq(objOwnIs(K), objOwnIs(k))\\
  \textbf{update(\_metadata)}{:}{-} eq(objOwnIs(K), objOwnIs(k))}\\
 \hline
 \makecell[l]{\emph{\#17 Right to be forgotten}}  & \makecell[l]{\textbf{delete}{:}{-} eq(objOwnIs(K), objOwnIs(k))}\\
 \hline
 \makecell[l]{\emph{\#21 Right to object}}  & \makecell[l]{\textbf{update}{:}{-} \text{gt}(\text{objPurIs}($K$) \& \text{objPurIs}($k$),0)\\
\textbf{read}{:}{-} \text{eq}(\text{objPurIs}($K$) \& \text{objObjIs}($k$),0)}\\
 \hline
 \makecell[l]{\emph{\#22 Automated individual}\\ \emph{decision-making}}  &  \makecell[l]{Derived from policies in \#5 and \#21}\\
 \hline
 \makecell[l]{\emph{\#25 Data protection by }\\ \emph{design and default}}  &\makecell[l]{
 \textbf{read}{:}{-} \text{eq(objOwnIs}($K$), \text{objOwnIs}($k$)) $\vee$ \\ gt(objOwnIs($K$) || \text{objShareIs}($k$), 0) \\
 \textbf{update}{:}{-} \text{eq(objOwnIs($K$), objOwnIs($k$))} \\
 \textbf{delete}{:}{-} \text{eq(objOwnIs($K$), objOwnIs($k$))}}\\
 \hline
 \makecell[l]{\emph{\#28 Do not grant unlimited }\\ \emph{access to data}}  & \makecell[l]{Access control based on session keys \\and ownership verification}\\
 \hline
 \makecell[l]{\emph{\#30 Records of }\\ \emph{processing activity}}  & \makecell[l]{\textbf{read}{:}{-} \text{eq(monitor($k$), true)}\\
\textbf{update}{:}{-} \text{eq(monitor($k$), true)}\\
\textbf{delete}{:}{-} \text{eq(monitor($k$), true)}}\\
 \hline
 \makecell[l]{\emph{\#31 Cooperation with }\\ \emph{supervisory authorities}}  & \makecell[l]{\projectname{}'s mutual attestation protocol,\\and tamper-evident logging scheme}\\
 \hline
 \makecell[l]{\emph{\#32 Security of processing}}  & \makecell[l]{Encryption of data and metadata,\\and TEE security properties}\\
 \hline
 \makecell[l]{\emph{\#33 Personal data breach}}  & \makecell[l]{
 \textbf{read}{:}{-} eq($K$, $K_{reg}$) $\wedge$ query({getLogs}()) }\\
 \hline
\end{tabular}
\caption{\gdpr articles related to KVS~\cite{gdprbench} and their mapping to \projectname{}'s policy enforcement rules. 
$K$: the session key of the querying entity, $k$: the requested data object, $K_{reg}$: the session key of a regulatory authority.
}
\vspace{-2mm}
\label{tab:gdpr_articles}
\end{table}

We next present how \projectname{}'s policy predicates express \gdpr articles related to KVS~\cite{gdprbench} through an e-commerce scenario (\autoref{fig:gdpr-entities-colour}): customer Alice (data owner), RecommendationService and AnalyticsService (processors), and an online retailer (controller). Alice's default policy allows recommendations and orders, objects to marketing and analytics, sets 90-day retention, and shares her data with RecommendationService, which in turn declares its recommendation purpose in its  execution policy, as shown in \autoref{sec:prog_model}. 
\autoref{tab:gdpr_articles} summarizes \projectname{}'s enforcement rules.

\myparagraph{Purpose limitation (Article 5)}
When RecommendationService attempts to read Alice's shopping preferences, \projectname{} validates that the declared purpose ``recommendations'' is included in Alice's allowed purposes and does not conflict with her objections  and sharing is permitted. The access succeeds, and since Alice's monitoring flag is enabled, the operation is logged for auditing.
\begin{lstlisting}[style=customc,mathescape]
query(get("alice:preferences")) $\wedge$ sessionKey(recommender_key)
    $\wedge$ objPurIs(recommendations)
\end{lstlisting}

In contrast, when AnalyticsService attempts to access the same data with purpose ``analytics'', \projectname{} detects this in Alice's objections, denies access and logs the violation attempt. %
\begin{lstlisting}[style=customc,mathescape]
query(get("alice:preferences")) $\wedge$ sessionKey(analytics_key)
    $\wedge$ objPurIs(analytics)
\end{lstlisting}

\myparagraph{Right of access (Article 15)}
Alice exercises her right to retrieve all her personal data stored in the retailer's system. \projectname{} locates all entries where Alice is the owner, and returns them. %
\begin{lstlisting}[style=customc,mathescape]
query(getm("alice:","data"))$\wedge$sessionKey(alice_key)$\wedge$objOwnIs(alice)
\end{lstlisting}

\myparagraph{Right to object (Article 21)}
After receiving targeted recommendations, Alice decides she no longer wants this processing. She exercises her right to object by updating her metadata across all entries. \projectname{} updates the objections bitmap in all affected entries while preserving the underlying data values. 
\begin{lstlisting}[style=customc,mathescape]
query(putm("alice:")) $\wedge$ sessionKey(alice_key) $\wedge$ objOwnIs(alice)
    $\wedge$ objObjections(marketing,analytics,recommendations)
\end{lstlisting}

\myparagraph{Right to be forgotten (Article 17)}
Alice requests complete deletion of her personal data. \projectname{} validates her ownership, locates all her entries, executes the deletion, and creates a comprehensive audit log documenting the deletion operation.
\begin{lstlisting}[style=customc,mathescape]
query(deletem("alice:")) $\wedge$ sessionKey(alice_key) $\wedge$ objOwnIs(alice)
\end{lstlisting}

\myparagraph{Records of processing activities (Article 30)}
The DataProtectionAuthority audits processing activities for Alice's preferences. \projectname{} authenticates the authority, retrieves the logs, performs integrity and freshness checks, and returns a formatted audit trail. 
\begin{lstlisting}[style=customc,mathescape]
query(getLogs("alice:preferences")) $\wedge$ sessionKey(regulator_key)
\end{lstlisting}

\section{Security Analysis}
\label{sec:security}

\begin{table}[t]
\centering
\footnotesize
\begin{tabular}{l|l} 
\hline
\textbf{Attack vector} & \textbf{Mitigation} \\
\hline
\hline
\textbf{System level attacks}\\
\hline
~~Read the \projectname{}'s memory & CVM protection \\
~~Read user data in the query request  & Data encryption \\
~~Load forged \projectname{} & Attestation \\
~~DMA to the \projectname{}'s memory & CVM protection \\
~~Manipulate user request & Data encryption \\
~~Rollback KV data & Undetected (out of scope)\\
\hline
\textbf{\gdpr-specific attacks}\\
\hline
~~Access other KV instance's memory & CVM protection \\
~~Read the \gdpr logs & Log encryption, access control \\
~~Tamper with \gdpr logs & Detection via integrity checks \\
~~Rollback \gdpr logs & Detection via freshness checks \\
~~Read KV data & Data encryption \\
~~Tamper with KV data & Detection via integrity checks \\
~~Malicious request for access to KV data & Policy check, access control \\
~~Fake user identity & Mutual attestation (optional) \\
\hline
\end{tabular}
\caption{Attack vectors and \projectname{}'s mitigations.}
\vspace{-3mm}
\label{tab:security_analysis}
\end{table}

Our security analysis comprehensively details potential attack vectors and their mitigations (\autoref{tab:security_analysis}), including both system-level and \gdpr-specific threats. 
Further, we formally verify \projectname{}'s attestation and logging protocols using the Tamarin Prover~\cite{tamarin_paper, tamarin_site} under the Dolev-Yao~\cite{dolev_yao} attacker model, detailed in~\autoref{appendix:security-protocol-analysis}.

For system-level attacks, \projectname{} leverages CVM protection to prevent unauthorized memory access, including DMA. 
Data encryption and the respective integrity checks protect query data %
and prevent data manipulation during transit. The attestation mechanism allows the detection of loading forged \projectname{} instances, ensuring only legitimate deployments can process sensitive data.
Note that \projectname{} does not protect against rollback attacks on database data, as this kind of attack requires an intrusive design in the format that databases store their data (e.g., versioning), which contradicts the compatibility goal of \projectname{}.
To prevent time-manipulation attacks, \projectname{} uses secure time sources provided by CVMs, e.g., AMD's secureTSC~\cite{securetsc}. %

For \gdpr-specific attacks, CVM-provided isolation prevents access to other KV instances' memory, while log encryption and access control protect \gdpr logs from unauthorized access. Integrity checks detect tampering with both logs and database data, and freshness checks prevent rollback attacks on logs. Lastly, \gdpr policy checks and access control mechanisms prevent malicious requests for unauthorized database access, and mutual attestation can prevent fake user identity attacks when enabled.

\section{Evaluation}
We evaluate \projectname{} by analyzing its \gdpr compliance performance (\autoref{subsec:eval:performance}), storage efficiency (\autoref{subsec:eval:storage}), and I/O subsystem (\autoref{subsec:eval:cvm_io}).
\if 
We structure the evaluation of \projectname{} around two key dimensions that demonstrate how \projectname{} balances \gdpr compliance with system performance while providing strong security guarantees: \emph{(i)}~performance and \gdpr compliance (\autoref{subsec:eval:performance}), and \emph{(ii)}~\projectname{}'s storage efficiency (\autoref{subsec:eval:storage}). Additionally, we provide an I/O analysis (\autoref{subsec:eval:cvm_io}) that characterizes the baseline performance costs of CVMs, contextualizing \projectname{}'s overhead within the broader landscape of secure cloud deployments.
\fi

\myparagraph{Prototype}
We implement \projectname{}'s policy compiler in \texttt{Python} and runtime components in \texttt{C++}. The KV adapter supports Redis~\cite{redis} and RocksDB~\cite{rocksdb}, communicating via Unix domain sockets, and uses \texttt{OpenSSL}~\cite{openssl} for cryptographic operations. 
\projectname{}'s logging system features a lossless \texttt{zlib}~\cite{zlib-citation, zlib} compression layer, which supports various compression levels (0-9), balancing between CPU consumption and compression efficiency~\cite{understanding-zlib, zlib-impl-comparison, zlib-benchmarks}.

\myparagraph{Testbed}
We perform our experiments on an AMD SEV-SNP server with an AMD EPYC 7713P CPU (64 cores, hyperthreading disabled) and 1024 GB of DDR4 DRAM.%
The server runs NixOS 25.05 with an AMD SEV SNP-enabled Linux kernel (v6.16.11). CVMs run Ubuntu 22.04 OS with a Linux kernel version v6.5.0.

\myparagraph{Baseline and variants}
We conduct experiments with the variants described in \autoref{tab:variants}. 
``GDPR'' employs \projectname{}, ``Encryption'' includes data encryption, and ``CVM'' deploys the system in CVMs. 
{\em Importantly, with Native KVS (deployed outside a VM), we show the worst-case overheads for  \projectname{}.}

\begin{table}[t]
\centering
\footnotesize
\begin{tabular}{c || c | c | c }
 \hline
 \textbf{Variant} & \textbf{GDPR} & \textbf{Encryption} & \textbf{CVM} \\
 \hline
 Native KVS 
 \textbf{\textcolor{red}{(deployed w/o VM)}} & \xmark & \xmark & \xmark\\
 \hline
 Native \projectname{} (w/o Encr) & \cmark & \xmark & \xmark \\ 
 \hline
 Native \projectname{} & \cmark & \cmark & \xmark \\ 
 \hline
 CVM KVS & \xmark & \xmark & \cmark\\
 \hline
 \projectname{} (w/o Encr) & \cmark & \xmark & \cmark\\
 \hline
 \projectname{} & \cmark & \cmark & \cmark\\
 \hline
\end{tabular}
\caption{Benchmarking variants.}
\vspace{-5mm}
\label{tab:variants}
\end{table}

\begin{table}[t]
\centering
\footnotesize
\begin{tabular}{c|l|l|l} 
\hline
\textbf{} & \textbf{Workload} & \textbf{Access pattern} & \makecell[l]{\textbf{Request}\\ \textbf{distribution}}  \\
\hline
\hline
\multirow{5}{*}{\rotatebox{90}{\textbf{YCSB\cite{ycsb}}}} 
& A & 50\% Read, 50\% Update & Zipfian\\ \cline{2-4}
& B & 95\% Read, 5\% Update & Zipfian\\ \cline{2-4}
& C & 100\% Read & Zipfian\\ \cline{2-4}
& D & 95\% Read, 5\% Insert & Latest\\ \cline{2-4}
& F & 100\% Read-Modify-Write & Zipfian\\
\hline
\multirow{3}{*}{\rotatebox{90}{\textbf{\gdpr\cite{gdprbench}}}}
& Controller & \makecell[l]{50\% Insert, 25\% Update metadata,\\ 25\% Delete} & Uniform\\ \cline{2-4}
& Customer & \makecell[l]{20\% Read, 20\% Read metadata, 20\% Update,\\ 20\% Update metadata, 20\% Delete} & Zipfian\\ \cline{2-4}
& Processor & \makecell[l]{80\% Read, 20\% Read by metadata} & Uniform\\
\hline
\end{tabular}
\caption{Benchmarking workloads and their characteristics.}
\vspace{-2mm}
\label{tab:workloads}
\end{table}

\begin{figure*}[t]
\centering
\includegraphics[]
{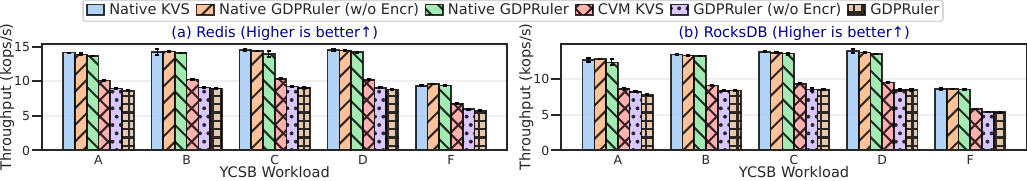}
\vspace{-1mm}
\caption{Workload analysis for Redis and RocksDB: \projectname{}'s performance with different YCSB workloads.}
\vspace{-2mm}
\label{fig:perf:throughput}
\end{figure*}

\begin{figure*}[t]
\centering
\includegraphics[]
{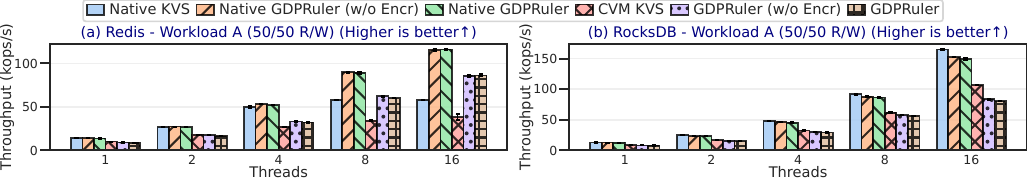}
\vspace{-1mm}
\caption{Scalability analysis for Redis and RocksDB: \projectname{}'s scalability with YCSB's write-heavy workload A. Other YCSB workloads are covered in
~\autoref{sec:appendix:additional_perf}.}
\vspace{-3mm}
\label{fig:perf:scalability}
\end{figure*}

\myparagraph{Workloads}
We use YCSB workloads (A-D, \& F)~\cite{Cooper2010, ycsb} to evaluate end-to-end KVS performance (we exclude workload E due to the lack of range queries in our prototype).
Each client uses a crafted default policy, controlling encryption and logging percentage. 

We also evaluate \projectname{} with \gdpr-specific workloads (\autoref{tab:workloads}) proposed by GDPRBench~\cite{gdprbench}. %
These workloads include: \emph{(i)}~\emph{Controller}, representing data controllers inserting KV pairs and performing administrative metadata updates and deletions, \emph{(ii)}~\emph{Customer}, representing data owners exercising \gdpr rights by reading their data, updating privacy preferences, and requesting deletions, and \emph{(iii)}~\emph{Processor}, representing data processors reading data with occasional metadata-based queries.%
We omit the \emph{Regulator} workload of GDPRBench as it was not publicly available by the authors.

In every run, we preload the KVS per YCSB guidelines~\cite{preload_ycsb}. Unless otherwise specified, we report the average across 3 runs.

\subsection{Performance under \gdpr Compliance}
\label{subsec:eval:performance}

We evaluate \projectname{} across three dimensions: \emph{(i)} KVS workload performance, \emph{(ii)} scalability, and \emph{(iii)} \gdpr workloads performance.

\myparagraph{KVS workload analysis}
For each YCSB workload, we execute 1M queries over 100K distinct keys. 
We set the logging to be 0\% to show the effect of processing the \gdpr metadata and performing their filtering in the hot data path.
For variants employing \projectname{}, we ensure all queries have valid policies, maintaining consistent data retrieval and insertion across experiments.

\autoref{fig:perf:throughput} presents the end-to-end throughput across all YCSB workloads with a single client (1 thread). %
In this setup, \projectname{} achieves 61.8\% and 61.7\% of the Native KVS throughput on average across all workloads for Redis and RocksDB, respectively.
The read-heavy Workload C reaches 61.4-62.3\% of baseline, while the respective range for the write-heavy Workload A is 61-62.1\%. The performance degradation mainly stems from write operations requiring metadata encoding and policy validation before KV insertion, while read operations primarily perform metadata extraction and filtering. 
We further observe that the deployment of the KVS engine in a CVM introduces 28.5-32.2\% overhead compared to native execution, mainly due to memory encryption and CVM I/O costs. Notably, adding \projectname{}'s \gdpr compliance on top of the CVM introduces only an additional 8.6\% overhead, while the encryption (AES-GCM) contributes minimally to this overhead ($<$2.6\%). %

\myparagraph{Scalability analysis}
To evaluate \projectname{}'s scalability, we vary the number of concurrent clients. \autoref{fig:perf:scalability} presents the scalability results for Redis and RocksDB for Workload A. 
\ifthenelse{\boolean{appendix}}
{Additional results for the rest YCSB workloads are provided in~\autoref{sec:appendix:additional_perf}.
}
{Additional results for the rest YCSB workloads are provided in Appendix D.
}

As thread count increases from 1 to 16, \projectname{} demonstrates good scalability with the throughput gap between the baseline and \projectname{} remaining relatively constant ($\approx$50\%). For instance, \projectname{} with RocksDB achieves 
80.6 K ops/sec for Workload A, while the baseline 
165.1 K ops/sec, when handling 16 concurrent clients.
However, for Redis deployments with more than 8 clients, \projectname{} using Unix sockets achieves comparable or better throughput than the native Redis with TCP connections.
Specifically, \projectname{} achieves 
86.1 K ops/sec for Workload A, while the baseline reaches only 
57.9 K ops/sec.
Our analysis indicates Redis's \emph{single-threaded} architecture creates a bottleneck at the TCP layer which \projectname{}'s connection handling bypasses. This phenomenon does not occur with our RocksDB server, which properly utilizes multiple threads and shows expected overhead patterns.

\begin{figure}[t]
\centering
\includegraphics[]
{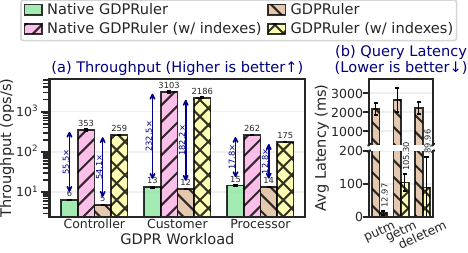}
\vspace{-2mm}
\caption{\projectname{}'s performance for \gdpr workloads and average latency of \gdpr queries with Redis backend. RocksDB results are covered in
~\autoref{sec:appendix:additional_perf}.}
\vspace{-1mm}
\label{fig:gdpr_workloads_perf}
\end{figure}

\myparagraph{\gdpr workloads}
We evaluate \projectname{}'s performance on \gdpr-specific workloads (Table~\ref{tab:workloads}) that use metadata-based operations (\emph{putm}, \emph{getm}, \emph{deletem}). We preload 100K KV pairs, then execute \gdpr workloads consisting of 1K operations with 8 connecting clients and encryption enabled, similarly to prior work~\cite{gdprbench}.
We run \projectname{} with (1) no indexing, and (2) ``user'' and ``purpose'' metadata indexing, which are the most frequently queried fields in \gdpr workloads.
We distribute KV pairs among 128 users and 128 data sources with round-robin assignment of purposes and objections. Each KV pair is configured to be shared with 10 users to reflect realistic multi-user access patterns and reduce query failure rates. We ensure that query operations specify valid purposes, maintaining consistency with realistic \gdpr-compliant access patterns.
\ifthenelse{\boolean{appendix}}
{We present the results of Redis here; RocksDB exhibits similar trends and is detailed in~\autoref{sec:appendix:additional_perf} due to space constraints.
}
{We present the results of Redis here; RocksDB exhibits similar trends and is detailed in Appendix D due to space constraints.
}

\autoref{fig:gdpr_workloads_perf}(a) demonstrates the throughput of \projectname{} on Controller, Customer, and Processor workloads for Redis. Without indexes, \projectname{} must perform full KVS scans to locate entries matching metadata criteria, resulting in significant performance drop as low as 4.8, 12.0, and 13.6 ops/sec  
for the three workloads respectively. This low throughput shows that even a relatively small subset of \gdpr queries---only 1K operations across 100K pairs---can become detrimental without proper indexing support. 
With \gdpr indexes enabled, performance improves by 12.8-182.2$\times$, depending on the workload. Specifically, Customer workloads that access all personal data of a specific owner benefit the most through indexed ownership lookups (182.2$\times$). Similarly, Processor workloads performing purpose-based filtering achieve significantly faster execution (12.8$\times$). 
These improvements highlight that \gdpr metadata indexing effectively eliminates the need for multiple full-KVS scans. 
Note that, similar to previous experiments, CVM deployment maintains 26.6-42.8\% overhead compared to native execution. %

\autoref{fig:gdpr_workloads_perf}(b) presents the latency for \gdpr operations averaged across the three workloads, justifying the poor performance in the non-indexed implementation.
\gdpr queries show significant latency reductions with indexing, from 2,178-2,654 ms to just 13-105 ms on average for Redis. Precisely, \emph{putm} operations improve by 168$\times$, while \emph{getm} operations achieve 25$\times$ speedup. This massive decrease is reasonable if we consider that, with indexes, operations transform from full KVS scans examining every KV pair to efficient index lookups followed by targeted KV pair fetching and metadata validation.
We further notice that even standard operations (\emph{put}, \emph{get}, \emph{delete}) benefit from reduced contention, showing 12-45\% latency improvements for Redis despite not directly using \gdpr indexes.

\subsection{Storage Efficiency}
\label{subsec:eval:storage}

We analyze the \projectname{}'s  storage layer characteristics across four dimensions: \emph{(i)}~the \emph{end-to-end impact} of logging on application throughput, \emph{(ii)}~the \emph{isolated throughput} of the logging subsystem itself, \emph{(iii)}~the \emph{storage overhead} introduced by embedding \gdpr metadata within KV records, and \emph{(iv)}~the \emph{audit log storage requirements} under varying logging configurations and compression settings.

\begin{figure}[t]
\centering
\includegraphics[]
{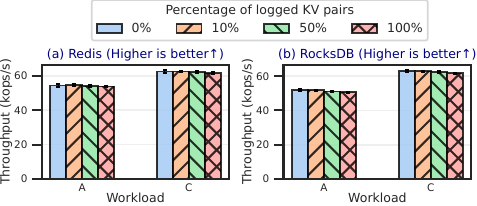}
\vspace{-1mm}
\caption{\projectname{}'s throughput with varying logged KV pair percentages for Redis and RocksDB.} %
\label{fig:performance_with_logging}
\end{figure}

\myparagraph{Logging layer impact on applications}
To quantify the impact of \gdpr audit logging, we run a write-heavy (Workload A) and a read-heavy (Workload C) workload with 8 client threads with varying logging percentages.
For these experiments, we configure \projectname{} with compression level 3 and encryption enabled.

\autoref{fig:performance_with_logging} highlights that \projectname{}'s asynchronous logging architecture incurs minimal performance overhead, with throughput degradation under 2\% even at 100\% logging across workloads and KVS. %
The low overhead stems from \projectname{}'s design decisions: asynchronous logging decouples it from the critical path, batched log writes amortize I/O costs across multiple entries, and efficient compression reduces the volume of data written to storage. These results show that \gdpr audit logging can be achieved in production systems without sacrificing performance, validating \projectname{}'s approach of treating logging as an off-critical-path operation.

\myparagraph{Logging layer performance}
To further understand \projectname{}'s logging system parameters, we evaluate the logging layer using a microbenchmark with 16 producer threads, varying logging threads (4 or 8), batch size (512-8192 entries), and log entry size (256 bytes, 1KB, 4KB), processing 10GB total in each configuration.
We use random-generated payloads in a format similar to the one used in YCSB workloads and disable compression to write an equivalent amount of data in each setup. 
We use Zipfian-distributed keys ($\theta$=0.99) across 100K files to simulate realistic patterns, with files written to a dedicated ext4 filesystem on a separate block device. 

\begin{figure}[t]
\centering
\includegraphics[]
{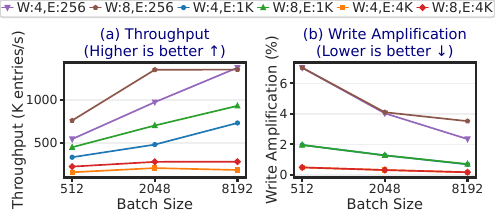}
\vspace{-1mm}
\caption{Logging layer performance analysis for various writer threads (W) and entry sizes (E).}
\vspace{-1mm}
\label{fig:logging-layer}
\end{figure}

\autoref{fig:logging-layer}(a) presents the throughput of \projectname{}'s logging system. We observe that batching effectiveness is inversely proportional to entry size: smaller entries (256 bytes) benefit significantly from batching, achieving substantial throughput gains as batch size increases, while larger entries (4KB) have diminishing returns. Further, the 8-writer configuration outperforms the 4-writer setup for most configurations, demonstrating effective parallelization. At batch size 2048, the 8-writer configuration achieves 1354, 705 and 283 K entries/sec for 256-byte, 1KB, and 4KB entries, respectively.

The throughput metric (K entries/s) reveals that the system achieves consistent and efficient I/O performance across different entry sizes, as larger entries inherently carry larger payloads. This validates that the logging system's I/O pipeline handles varying data volumes effectively. The results also demonstrate an interaction between CPU resources and batching: with 8 writers, the system reaches a plateau shown in batch sizes 2048 and 8192. In contrast, the resource-constrained configuration (4 threads) continues to benefit from larger batch sizes, as it compensates for limited CPU parallelism by amortizing per-batch processing overhead.

\autoref{fig:logging-layer}(b) shows write amplification, measuring the ratio of actual bytes written to storage versus the original payload size. Write amplification remains consistently low (below 7\%) across all configurations, validating that \projectname{}'s logging format introduces minimal metadata overhead. The slight increase in write amplification for smaller entry sizes (256 bytes showing ~6-7\% overhead) reflects the relatively higher cost of per-entry headers, while larger entries (4KB showing $<1\%$ overhead) amortize this cost.

\begin{table}[t]
\centering
\setlength{\tabcolsep}{3pt} %
\footnotesize
\begin{tabular}{lcccccc}

\toprule
\multirow{2}{*}{\textbf{KVS}} & \multirow{2}{*}{\parbox{1.0cm}{\centering\textbf{KVS (MB)}}} & \multirow{2}{*}{\parbox{1.3cm}{\centering\textbf{\projectname{} KVS (MB)}}} & \multirow{2}{*}{\parbox{0.8cm}{\centering\textbf{Compr.\\level}}} & \multicolumn{3}{c}{\textbf{\gdpr logs (MB)}} \\
\cmidrule{5-7}
& & & & \textbf{10\%} & \textbf{50\%} & \textbf{100\%} \\
\midrule

\multirow{3}{*}{\textbf{Redis}} 
& \multirow{3}{*}{103.27} & \multirow{3}{*}{112.42} 
& 0 & 44.47 & 221.11 & 463.20 \\
& & & 3 & 7.18 & 43.02 & 87.97 \\
& & & 6 & 6.80 & 41.62 & 84.45 \\

\midrule

\multirow{3}{*}{\textbf{RocksDB}} 
& \multirow{3}{*}{144.40} & \multirow{3}{*}{173.06} 
& 0 & 44.44 & 221.07 & 463.17 \\
& & & 3 & 7.02 & 43.03 & 88.03\\
& & & 6 & 6.80 & 41.63 & 84.44 \\

\bottomrule
\end{tabular}
\caption{Storage requirements analysis for \projectname{}.}
\vspace{-3mm}
\label{tab:log_write_amplification}
\end{table}

\myparagraph{KVS metadata storage overhead}
Table~\ref{tab:log_write_amplification} quantifies the storage overhead introduced by \gdpr metadata. %
Redis KVS consumes 103.27 MB for 100K KV pairs with 1KB values, while \projectname{} requires 112.42 MB for KVS files, an 8.9\% increase. RocksDB shows a similar pattern with 19.8\% metadata overhead (144.40 MB vs. 173.06 MB). This overhead stems from embedding compliance metadata---purposes, objections, ownership, retention periods, and sharing permissions---in each value and is inversely proportional to value size; larger values dilute the relative metadata cost.

\myparagraph{Audit log storage requirements}
Table~\ref{tab:log_write_amplification} presents \gdpr log storage requirements with varying log percentages and compression levels. Without compression (level 0), logging all operations (100\%) generates 463.20 MB of logs for 100K entries with 1M operations, yielding a 4.3:1 ratio of logs to original KVS size (103.27 MB). This substantial volume reflects comprehensive audit trail requirements where each operation generates a metadata-rich log entry. At 50\% logging—representing realistic selective monitoring involving only sensitive data—uncompressed logs consume 221.11 MB (almost twice the KVS size), while conservative 10\% logging generates 44.47 MB of audit trail. 
However, when compression (level 3) is used, it reduces log storage by 5$\times$ overall, reaching $\approx$88 MB for 100\% logging. At 50\%, logs are 43 MB ($\approx$40\% of KVS size). 
Level 6 offers negligible extra savings ($\approx$2.1\%) but increases CPU overhead~\cite{understanding-zlib, zlib-impl-comparison, zlib-benchmarks}.

\subsection{Data I/O Subsystem Analysis}
\label{subsec:eval:cvm_io}
To contextualize \projectname{}'s performance results and isolate the baseline costs of confidential computing from \gdpr compliance, we evaluate the I/O behavior of CVMs.

\begin{figure}[t]
\centering
\includegraphics[]
{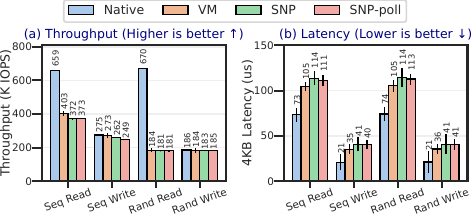}
\vspace{-1mm}
\caption{FIO performance analysis.}
\vspace{-1mm}
\label{fig:cvm_storage}
\end{figure}

\myparagraph{Storage I/O performance}
\autoref{fig:cvm_storage} shows FIO benchmark~\cite{fio} results comparing 4KB IOPS across native execution, standard VM, and CVM (SNP) configurations. We observe that sequential reads degrade to 61\% and 57\% of the native variant while random reads drop even further, to just 27\% of native throughput for both VM and CVM configurations. In contrast, both sequential and random writes maintain near-native performance (95-99\% of the native variant) for both VM and CVM configurations.
We further observe that enabling halt polling~\cite{guest-halt-poll} (SNP-poll) does not result in any significant differences from standard SNP configurations across all workloads. The cost of memory encryption remains modest (8-15\%), demonstrating that virtualization dominates performance impact. These measurements align with previous CVM characterizations~\cite{damon_price_of_privacy, cvm-eval}. Importantly, \projectname{}'s logging system operates in sequential write patterns where CVMs maintain 95\% of native performance, validating the almost negligible <2\% logging overhead (\autoref{subsec:eval:storage}).

\begin{figure}[t]
\centering
\includegraphics[]
{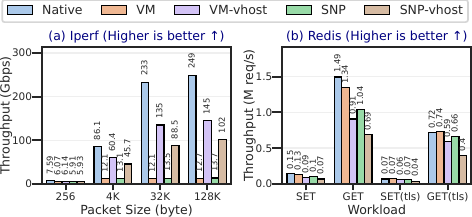}
\vspace{-1mm}
\caption{Iperf and Redis performance (multiqueue=8).}
\vspace{-1mm}
\label{fig:cvm_network}
\end{figure}

\myparagraph{Network I/O characteristics}
\autoref{fig:cvm_network} presents network performance evaluation %
using iperf3~\cite{iperf} (a) and Redis (b) with a virtio-net device (8 queues, 8 kernel vhost threads on host, 8 interrupt vectors in guest).
vhost optimization~\cite{vhost-net} (VM-vhost, SNP-vhost) significantly improves throughput for large messages: for 128KB transfers, VM-vhost achieves 145 Gbps and SNP-vhost reaches 102 Gbps, compared to 12-14 Gbps for standard VM/SNP configurations. %
However, even so, CVMs show up to 60\% performance drop compared to native TCP, particularly as packet size increases. 
Redis workloads exhibit opposite behavior. For GET queries, standard VM achieves 1.34 M req/s while VM-vhost delivers only 0.91 M req/s and SNP-vhost 0.69 M req/s. While vhost with multiqueue excels at iperf's bulk transfers via parallel processing, Redis's request-response pattern generates excessive software interrupt requests (softIRQs) and context switches from multiple host threads.
This causes interrupt overhead that dominates parallelization benefits, making standard paravirtualization a better choice for KVS workloads.

\section{Related work}

\myparagraph{Policy compliance}
A significant line of work explores policy-based compliance systems to address the requirements of privacy regulations. 
Systems such as Qapla~\cite{qapla}, K9db~\cite{k9db}, and Deep Enforcement~\cite{deep_enforcement} focus on policy-driven schema transformations and access control to ensure policy compliance. 
Other efforts, mainly for relational databases, formalize and enforce data policies using policy languages~\cite{blindseer, picachv, sieve, ifm2022, legalease,ironsafe}, optimize for performance~\cite{shai}, or propose storage-centric policy compliance solutions~\cite {guardat, sw_defined_data_protection, pesos}.
On the \gdpr frontier, prior work examines compliance challenges in the cloud~\cite{mohan2019analyzinggdprcompliancelens, gdpr_anti_patterns, sins_of_gdpr,gdprxiv}, its impact on storage systems~\cite{gdpr_on_storage_systems}, and proposed techniques to enforce \gdpr regulations~\cite{position_gdpr_by_construction,datacase, gdprbench}. 
Unlike these approaches, \projectname{} is an end-to-end system for \gdpr compliance that couples a rich policy language with verifiable execution for KVS.

\myparagraph{Compliance plugins}
Prior studies that investigate compliance frameworks for existing systems either provide transparent runtime policy enforcement for legacy SQL databases~\cite{gdprizer, fontus,shengendb}, enforce policies at the programming language layer~\cite{dpl}, develop specialized auditing frameworks~\cite{hippocratic}, or design enforcement systems for web applications~\cite{rulekeeper, edna, blockaid} or storage systems~\cite{pesos, thoth, ironsafe}.
These solutions either require intrusive application changes, such as at the database layer~\cite{gdprbench, k9db}, placing the burden on developers to build compliance mechanisms~\cite{gdprizer}, or lack the security guarantees needed for compliance~\cite{rulekeeper}.
Moreover, they fail to provide such compliance in a verifiable manner on untrusted cloud infrastructures.
Compared to them, \projectname{} enables \gdpr compliance for KVS without requiring changes to the application or KVS engines. %

\myparagraph{Secure data management systems}
A line of work focuses on secure cloud database and storage systems, including systems that ensure data security at the database level~\cite{arx,concerto} and support relational queries over encrypted data~\cite{monomi,cryptdb,hedb,azure_always_encrypted,cipherbase}.
Hardware-based solutions~\cite{enclavedb, speicher, anchor, pldb, trusteddb, veridb, precursor, enclage, lsm_fast22, obliviate, StrongBox} leverage TEEs to protect data during processing and at rest, while others extend security guarantees to distributed cloud environments~\cite{treaty, avocado}.
In contrast, \projectname{} focuses on \gdpr compliance, addressing purpose limitation, data subject rights, and auditable logging, while preserving strong security guarantees for both data and metadata and ensuring compatibility with existing KV store deployments.

\section{Conclusion}
\label{sec:conclusion}
In this paper, we present \projectname{}, an end-to-end system enabling \gdpr compliance for KVS without codebase modification. \projectname{} balances compliance and system performance using a comprehensive policy language, a secure runtime monitor, and an efficient metadata management system. Deploying the enforcement engine within a CVM ensures private data and metadata protection throughout the processing pipeline. Our evaluation shows that \projectname{} achieves \gdpr compliance with reasonable performance overhead, minimizes metadata storage via optimized encoding, and provides tamper-evident logging and efficient \gdpr queries, demonstrating its suitability for cloud adoption.

\subsection*{Acknowledgments}
This work was supported in part by an ERC Starting Grant (ID: 101077577) and the Chips Joint Undertaking (JU), European Union (EU) HORIZON-JU-IA, under grant agreement No. 101140087\linebreak (SMARTY), the Intel Trustworthy Data Center of the Future (TDCoF), Google Research Grants,  Portuguese national funds through Funda\c{c}\~ao para a Ci\^encia e a Tecnologia, I.P. (FCT) under projects UID/50021/2025 (DOI: \url{https://doi.org/10.54499/UID/50021/2025}) and UID/PRR/50021/2025 (DOI: \url{https://doi.org/10.54499/UID/PRR/50021/2025}), IAPMEI under grant C6632206063-00466847 (PT Smart Retail), and the European Union’s Horizon Europe Research and Innovation Programme under Grant Agreement No. 101189689.

\section*{Ethical Considerations}

The research team attests that we considered the ethics of this research.
All experiments and evaluations were executed exclusively on local computing environments, ensuring that no live systems were affected.
The research did not require the collection of personal or sensitive data.
The findings of this research are intended to establish \gdpr compliance and ensure data privacy and security for data management systems deployed on untrusted cloud infrastructures, and are intended for constructive and beneficial purposes.

\section*{Open Science}
We openly share our research artifact, including the \projectname{} framework (\href{https://github.com/dimstav23/GDPRuler}{{\color{blue}Github link}}) and its associated security protocol proofs (\href{https://github.com/dimstav23/GDPRuler-proofs}{{\color{blue}Github link}}). 
Our artifact includes \projectname{}'s code, its policy language and compiler, its formal verification proofs, and is accompanied by the respective README files. 
It also incorporates scripts that generate the workloads used in \projectname{}'s evaluation and others that provide guidance for its configuration, deployment, and evaluation on AMD SEV-SNP machines.

\balance
\bibliographystyle{plain}
\bibliography{main}

\ifthenelse{\boolean{appendix}}
{
\clearpage
\appendix
\newpage
\appendix
\section*{Appendix}
\label{appendix}

We provide supplementary material for \projectname{} that includes: 
\begin{enumerate}

\item {\bf Detailed algorithms} for \projectname{}'s \gdpr-aware operations (\autoref{sec:appendix:algos}).

\item {\bf Formal verification proofs} of \projectname{}'s attestation and logging protocols (\autoref{appendix:security-protocol-analysis}).

\item {\bf Additional experimental results} on \projectname{}'s scalability, \gdpr queries performance and compression efficiency (\autoref{sec:appendix:additional_perf}).

\item {\bf Legislation framework analysis} which confirms that the \gdpr requirements, enforced in \projectname{}, align with the core body of other major regulations, showcasing \projectname{}'s adaptability to other/future privacy legislations (\autoref{tab:privacy_laws_comparison} in \autoref{sec:appendix:legislation}).
\end{enumerate}

\section{Detailed \gdpr-Aware Operations}
\label{sec:appendix:algos}

{\LinesNumberedHidden
\setlength{\algomargin}{0pt}
\setlength{\parskip}{0pt}
\begin{algorithm}[ht]
\myfontsizealgorithms
\SetAlgoLined
\SetKwInOut{KwIn}{Input}
\SetKwInOut{KwOut}{Output}
\KwIn{~key, session\_key, query\_policy}
\KwOut{~Value or \gdpr violation error}
\underline{{\bf get($key$)}} \\
\Begin{
    /* \texttt{Policy compilation and validation} */\\
    $default\_policy \leftarrow$ \texttt{get\_session\_policy}($session\_key$);\\    
    $policy \leftarrow$ \texttt{compile\_query\_policy}($default\_policy$, $query\_policy$);
    
    /* \texttt{Standard KV operation} */\\
    $value \leftarrow$ \texttt{decrypt}(\texttt{kv\_store.get}($key$);\\
    \If{$value$ == NULL}{
        \Return \texttt{GET\_FAILED};\\
    }
    
    /* \texttt{Metadata extraction and validation} */\\
    $gdpr\_metadata \leftarrow$ \texttt{extract\_metadata}($value$);\\
    $compliance \leftarrow$ \texttt{validate}($gdpr\_metadata$, $policy$, $session\_key$);\\
    
    \If{$\neg compliance$}{
        /* \texttt{Log failed access attempt} */\\
        \texttt{monitor\_query}($key$, \texttt{``get''}, $compliance$);\\
        \Return \texttt{GDPR\_VIOLATION\_ERROR};\\
    }
    
    /* \texttt{Audit logging if required} */\\
    \If{$gdpr\_metadata$.\texttt{monitor} == $true$}{
        \texttt{monitor\_query}($key$, \texttt{``get''}, $compliance$);\\
    }
    
    /* \texttt{Return processed value} */\\
    \Return \texttt{remove\_gdpr\_metadata}($value$);\\
}
\caption{GDPR-Aware Get Operation}
\label{alg:gdpr-get}
\end{algorithm}
}

\subsection{GET Operation}
The \gdpr-aware get operation (\autoref{alg:gdpr-get}) demonstrates how \projectname{} transforms simple data retrieval into a comprehensive compliance-checking process. The operation begins by compiling the requesting entity's session policy with any query-specific overrides, creating a unified policy framework that captures both default privacy preferences and operation-specific requirements. After retrieving the encrypted value from the underlying KVS and performing decryption within the secure CVM environment, the system extracts embedded \gdpr metadata (e.g., ownership information, processing purposes, retention policies, sharing permissions).

The core compliance validation ensures that the requesting entity has legitimate access rights based on \gdpr principles: ownership verification (Article 15), purpose limitation checking (Article 5), and objection validation (Article 21). If validation fails, the system logs the unauthorized access attempt for audit purposes and returns a \gdpr violation error, ensuring that privacy breaches are both prevented and documented. For successful operations, audit logging occurs conditionally based on the data's monitoring requirements, creating tamper-proof records that regulatory authorities can later review. The final step removes \gdpr metadata from the returned value, ensuring that clients receive only the actual data while the compliance infrastructure remains transparent.

{\LinesNumberedHidden
\setlength{\algomargin}{0pt}
\begin{algorithm}[ht]
\myfontsizealgorithms
\SetAlgoLined
\SetKwInOut{KwIn}{Input}
\SetKwInOut{KwOut}{Output}
\KwIn{~key, value, session\_key, query\_policy}
\KwOut{~Success or failure status}
\underline{{\bf put($key, value$)}} \\
\Begin{
    /* \texttt{Policy compilation and validation} */\\
    $default\_policy \leftarrow$ \texttt{get\_session\_policy}($session\_key$);\\
    $policy \leftarrow$ \texttt{compile\_query\_policy}($default\_policy$, $query\_policy$);
    
    /* \texttt{Check metadata cache, fallback to KVS if needed} */\\
    $gdpr\_metadata \leftarrow$ \texttt{metadata\_cache.get}($key$);\\
    \If{$gdpr\_metadata$ == NULL}{
        /* \texttt{Cache miss: retrieve from KV store} */\\
        $value \leftarrow$ \texttt{decrypt}(\texttt{kv\_store.get}($key$));\\
        \If{$value$ != NULL}{
            $gdpr\_metadata \leftarrow$ \texttt{extract\_metadata}($value$);\\
            \texttt{metadata\_cache.put}($key$, $gdpr\_metadata$); // \texttt{Update cache}\\
        }
    }
    /* \texttt{Validate modification permissions for existing keys} */\\
    \If{$gdpr\_metadata$ != NULL}{
        $compliance \leftarrow$ \texttt{validate}($gdpr\_metadata$, $policy$, $session\_key$);\\
        \If{$\neg compliance$}{
            \texttt{monitor\_query}($key$, \texttt{``put''}, $compliance$);\\
            \Return \texttt{PUT\_FAILED};\\
        }
    }
    
    /* \texttt{Metadata creation/update via query rewriter} */\\
    $enhanced\_value \leftarrow$ $rewriter$.\texttt{create\_new\_value}($policy$, $value$);\\
    
    /* \texttt{Audit logging if required} */\\
    \If{$policy$.\texttt{monitor} == $true$}{
        \texttt{monitor\_query}($key$, \texttt{``put''}, $enhanced\_value$, $compliance$);\\
    }

    /* \texttt{Encryption if required} */\\
    \If{$policy$.\texttt{encrypt} == $true$}{
        $enhanced\_value \leftarrow$ \texttt{encrypt}($enhanced\_value$);\\
    }
    
    /* \texttt{Standard KV operation with enhanced value} */\\
    $result \leftarrow$ \texttt{kv\_store.put}($key$, ($enhanced\_value$));\\
    \Return $result$ ? \texttt{PUT\_SUCCESS} : \texttt{PUT\_FAILED};\\
}
\caption{GDPR-Aware Put Operation}
\label{alg:gdpr-put}
\end{algorithm}
}

\subsection{PUT Operation}
The \gdpr-aware put operation (\autoref{alg:gdpr-put}) extends standard KV storage with comprehensive privacy policy enforcement and metadata management. After compiling session and query-specific policies, the operation checks whether the key already exists by first consulting an in-memory metadata cache. This cache optimization reduces KVS load by avoiding repeated retrievals of frequently accessed metadata. On cache miss, the system retrieves and decrypts the value from the KVS, extracts its metadata, and updates the cache for future requests. For existing keys, the system validates that the requesting entity has modification rights by checking ownership against the cached or retrieved metadata, preventing unauthorized alterations to protected data.

The query rewriter component creates an enhanced value that embeds \gdpr metadata directly within the stored data, encoding privacy policies, ownership information, processing purposes, and retention settings using efficient metadata representations. This embedded approach ensures that compliance information travels with the data throughout its lifecycle while maintaining compatibility with existing KVS implementations. The system performs conditional audit logging based on policy requirements, then encrypts and stores the enhanced value. This approach ensures that every write operation respects existing privacy preferences while establishing comprehensive governance for new data entries.

{\LinesNumberedHidden
\setlength{\algomargin}{0pt}
\begin{algorithm}[ht]
\myfontsizealgorithms
\SetAlgoLined
\SetKwInOut{KwIn}{Input}
\SetKwInOut{KwOut}{Output}
\KwIn{~key\_prefix, filter\_conditions}
\KwOut{~Filtered results or failure}
\underline{{\bf getm($key\_prefix, filter\_conditions$)}} \\
\Begin{
    /* \texttt{Bulk retrieval from KV store} */\\    
    /* \texttt{Query metadata indexes for matching keys} */\\
    $candidate\_keys \leftarrow$ \texttt{query\_metadata\_indexes}($filter\_conditions$);\\
    
    \If{$candidate\_keys$.empty()}{
        /* \texttt{Fallback to full KVS scan if no index available} */\\
        $candidate\_keys \leftarrow$ \texttt{get\_all\_keys}($key\_prefix*$);\\
    }
    $encrypted\_values \leftarrow$ \texttt{kv\_store.get}($candidate\_keys$);\\
    $values \leftarrow$ \texttt{decrypt($encrypted\_values$});\\
    \If{$values$.\texttt{empty}()}{
        \Return \texttt{GETM\_FAILED};\\
    }
        
    /* \texttt{Filter and validate each KV pair} */\\
    $has\_results \leftarrow$ false;\\
    \ForEach{$value \in values$}{
        $gdpr\_metadata \leftarrow$ \texttt{extract\_metadata}($value$);\\
        $gdpr\_filter \leftarrow$ \texttt{create\_gdpr\_filter}($gdpr\_metadata$);\\
        
        /* \texttt{Apply filter conditions} */\\
        \If{$\neg$ \texttt{apply\_filter}($gdpr\_filter$, $filter\_conditions$)}{
            \texttt{continue}; // \texttt{Skip non-matching filter entries}\\
        }
        
        /* \texttt{Validate access permissions} */\\
        $compliance \leftarrow$ \texttt{validate}($gdpr\_metadata$, $policy$, $session\_key$);\\
        \texttt{monitor\_query}($key$, \texttt{``getm''}, $compliance$);\\
        
        \If{$compliance$}{
            $compliant\_entries$.\texttt{append}(\texttt{remove\_gdpr\_metadata}($value$));\\
            $has\_results \leftarrow$ true;\\
        }
    }
    
    \Return $has\_results$ ? $compliant\_entries$ : \texttt{GETM\_FAILED};\\
}
\caption{GDPR-Aware Getm Operation}
\label{alg:gdpr-getm}
\end{algorithm}
}

\subsection{DELETE Operation}
The \gdpr-aware delete operation performs the same process as the put but instead of invoking the query rewriter, it submits the request for the deletion of a KV pair directly to the KV engine, after ensuring that the requesting entity has modification rights (ownership and access permissions checks), preventing unauthorized deletions of protected data.

\subsection{GETM Operation}
The \gdpr-aware getm operation (\autoref{alg:gdpr-getm}) enables efficient bulk retrieval with privacy-aware filtering, essential for implementing \gdpr rights like "right of access" or "data portability". After policy compilation, the system performs optimized data retrieval. For improved performance, \projectname{} first queries metadata indexes to identify candidate keys matching the specified filter conditions (e.g., owner, purposes, expiration times). If indexes are available, this dramatically reduces the search space by eliminating non-matching entries early; otherwise, the system falls back to key prefix matching for bulk retrieval. After decryption and metadata extraction, the operation applies a two-stage filtering process: \gdpr metadata filters select entries matching specified criteria, followed by individual access validation ensuring the requesting entity has legitimate rights to access each matching entry

This dual filtering approach, combined with metadata indexing, optimizes performance by eliminating unauthorized entries early while maintaining strict compliance validation. The operation aggregates all qualifying results and logs each successful access individually for audit purposes, creating detailed trails that show exactly which data was accessed and why. This batch processing capability with index support is crucial for large-scale privacy operations while ensuring that each data item receives individual compliance validation.

\subsection{PUTM Operation}
The \gdpr-aware putm operation (\autoref{alg:gdpr-putm}) provides bulk metadata updates essential for implementing policy changes across multiple data entries. After policy compilation, the operation uses index queries to efficiently locate entries matching filter conditions before performing bulk retrieval and decryption. This index-accelerated approach significantly reduces the performance overhead of bulk metadata operations compared to full KVS scans.

The system applies \gdpr metadata filters to the retrieved entries and validates modification permissions individually, ensuring that policy changes respect ownership boundaries. The query rewriter creates updated values by modifying metadata fields while preserving the underlying data values, ensuring that privacy policy changes don't affect data integrity. Each successful modification is logged, if required by the policy, for audit purposes, providing complete traceability for bulk policy operations. The bulk update operation ensures consistency across all modifications. This approach enables efficient implementation of complex \gdpr operations.
{\LinesNumberedHidden
\setlength{\algomargin}{0pt}
\begin{algorithm}[h]
\myfontsizealgorithms
\SetAlgoLined
\SetKwInOut{KwIn}{Input}
\SetKwInOut{KwOut}{Output}
\KwIn{~key\_prefix, filter\_conditions, new\_metadata}
\KwOut{~Update status}
\underline{{\bf putm($key\_prefix, filter\_conditions, new\_metadata$)}} \\
\Begin{
    /* \texttt{Bulk retrieval from KV store} */\\
    /* \texttt{Query metadata indexes for matching keys} */\\
    $candidate\_keys \leftarrow$ \texttt{query\_metadata\_indexes}($filter\_conditions$);\\
    
    \If{$candidate\_keys$.empty()}{
        /* \texttt{Fallback to full KVS scan if no index available} */\\
        $candidate\_keys \leftarrow$ \texttt{get\_all\_keys}($key\_prefix*$);\\
    }

    $encrypted\_values \leftarrow$ \texttt{kv\_store.get}($candidate\_keys$);\\  
    \If{$values$.\texttt{empty}()}{
        \Return \texttt{PUTM\_FAILED};\\
    }
    
    /* \texttt{Filter, validate and prepare updates} */\\
    $failed\_count \leftarrow$ 0;\\
    \ForEach{$(key, current\_value) \in (candidate\_keys, values)$}{
        $current\_value \leftarrow$ \texttt{decrypt}($current\_value$);\\  
        $gdpr\_metadata \leftarrow$ \texttt{extract\_metadata}($current\_value$);\\
        $gdpr\_filter \leftarrow$ \texttt{create\_gdpr\_filter}($metadata$);\\
        
        /* \texttt{Apply filter conditions} */\\
        \If{$\neg$\texttt{apply\_filter}($gdpr\_filter$, $filter\_conditions$)}{
            \texttt{continue}; // \texttt{Skip non-matching filter entries}\\
        }
        
        /* \texttt{Validate modification permissions} */\\
        $compliance \leftarrow$ \texttt{validate}($gdpr\_metadata$, $policy$, $session\_key$);\\
        
        \uIf{$compliance$}{
            /* \texttt{Create updated value with new metadata} */\\
            $new\_value \leftarrow$ $rewriter$.\texttt{create\_new\_value}($current\_value$, $new\_metadata$);\\
            \texttt{monitor\_query}($key$, \texttt{``putm''}, $compliance$, $new\_value$);\\
            /* \texttt{Encryption if required} */\\
            \If{$new\_metadata$.\texttt{encrypt} == $true$}{
                $new\_value \leftarrow$ \texttt{encrypt}($new\_value$);\\
            }
            $valid\_updates$.\texttt{append}($key, new\_value$);\\
        }
        \Else{
            $failed\_count \leftarrow$ $failed\_count$ + 1;\\
        }
    }
    
    /* \texttt{Bulk update operation} */\\
    $update\_results \leftarrow$ \texttt{kv\_store.put}($valid\_updates$);\\
    $updated\_count \leftarrow$ \texttt{count\_successes}($update\_results$);\\
    
    \Return $updated\_count$ > 0 ? \texttt{PUTM\_SUCCESS} : \texttt{PUTM\_FAILED};\\
}
\caption{GDPR-Aware Putm Operation}  
\label{alg:gdpr-putm}
\end{algorithm}
}

\subsection{DELETEM Operation}
The \gdpr-aware deletem operation performs a similar process with the PUTM. However, instead of creating new values, it locates and filters the compliant KV pairs that match the filter conditions and submit deletion requests directly to the KVS while also respecting the KV pairs' logging requirements. 

{\LinesNumberedHidden
\setlength{\algomargin}{0pt}
\begin{algorithm}[t]
\myfontsizealgorithms
\SetAlgoLined
\SetKwInOut{KwIn}{Input}
\SetKwInOut{KwOut}{Output}
\KwIn{~key\_filter, timestamp\_threshold, session\_key}
\KwOut{~Audit log entries or access denied}
\underline{{\bf getLogs($key\_filter, time\_thres, session\_key$)}} \\
\Begin{
    /* \texttt{Validate regulator authority} */\\
    $regulator\_valid \leftarrow$ \texttt{validate\_regulator\_key}($session\_key$);\\
    \If{$\neg regulator\_valid$}{
        \Return \texttt{ACCESS\_DENIED};\\
    }
    
    /* \texttt{Briefly pause logging to ensure consistency} */\\
    \texttt{pause\_logger\_and\_flush\_logs}();\\
    
    /* \texttt{Retrieve filtered audit logs} */\\
    \uIf{$key\_filter$ != NULL}{
        /* \texttt{Get logs for specific key} */\\
        /* \texttt{read\_log: decryption, integrity and counter checks} */\\
        $total\_log\_entries \leftarrow$ \texttt{read\_log}($hash(key\_filter)$, $time\_thres$);\\
    }
    \Else{
        /* \texttt{Get all available log files} */\\
        $log\_files \leftarrow$ \texttt{retrieve\_log\_names}();\\
        \ForEach{$log\_file \in log\_files$}{
            /* \texttt{read\_log: decryption, integrity and counter checks} */\\
            $entries \leftarrow$ \texttt{read\_log}($log\_file$, $time\_thres$);\\
            $total\_log\_entries$.\texttt{append}($entries$);\
        }
    }
    
    /* \texttt{Format logs for regulatory review} */\\
    \ForEach{$entry \in total\_log\_entries$}{
        $readable\_entry \leftarrow$ \texttt{gdpr\_format}($entry$);\\
        $formatted\_logs$.\texttt{append}($readable\_entry$);\
    }
    
    \Return $formatted\_logs$;\\
}
\caption{GDPR Audit Log Access for Regulators}
\label{alg:gdpr-getlogs}
\end{algorithm}

\subsection{GetLogs Operation}
The \gdpr-aware getLogs operation (\autoref{alg:gdpr-getlogs}) provides regulatory authorities with secure access to comprehensive audit trails essential for compliance verification and violation investigation. The operation begins with strict authentication to verify that only authorized regulatory entities can access sensitive audit information. The system temporarily pauses ongoing logging operations to ensure consistency in the retrieved audit trail, preventing data races that could compromise audit integrity.

The flexible retrieval mechanism supports both specific key-based queries and comprehensive audit reviews, with timestamp filtering enabling investigation of specific time periods. Each log entry undergoes decryption, integrity verification, and counter validation to ensure authenticity and detect any tampering attempts. The final formatting phase converts technical log entries into human-readable reports that include complete \gdpr metadata interpretation, operation histories, and compliance decisions. This comprehensive audit capability enables regulatory authorities to verify compliance, investigate potential violations, and understand the complete lifecycle of personal data processing within the system.

\section{Security Protocol Analysis}\label{appendix:security-protocol-analysis}

We formally verify \projectname{}'s attestation and logging protocols using the Tamarin Prover~\cite{tamarin_paper, tamarin_site} under the Dolev-Yao~\cite{dolev_yao} attacker model. We prove that the attestation protocol successfully performs mutual authentication and allows the creation of a secure direct communication channel between any client and the \gdpr monitor. Additionally, we prove the correctness of the logging protocol by guaranteeing that after successful log validation, the log is complete with only valid entries. 

\subsection{Threat Model}
Tamarin natively incorporates a Dolev-Yao~\cite{dolev_yao} attacker model for the network, which we extend with additional attacker capabilities depending on the protocol we analyze. The Dolev-Yao attacker model assumes a strong attacker with full control over the network, who is able to read, modify, and drop any messages. For the logging protocol, we strengthen the attacker to be able to delete, modify, and add arbitrary log entries directly in untrusted storage. We further consider how an attacker capable of compromising the long-term keys of the \gdpr monitor or any client affects the security properties we verify. For our models, we assume that any cryptographic primitives (e.g., hashing, signatures) are perfect and cannot be attacked.

\subsection{Protocol Model}
We model our protocols as multiset rewriting rules, one of the supported input formats of the Tamarin Prover. The global system state corresponds to a multiset of facts, where a fact represents some piece of information like a network message, the state of the \gdpr monitor, attacker knowledge, etc. Next, we explain each protocol's model in more detail:

\myparagraph{Attestation model}
Our attestation model assumes the TEE attestation infrastructure to work correctly, i.e., it does not leak secrets and can not be attacked. Two key properties are shown for the attestation protocol: \emph{(i)}~\textbf{authenticity}, after attestation, the \gdpr monitor can trust the identity of the entity making the request and vice-versa; and \emph{(ii)}~\textbf{perfect-forward-secrecy}, the symmetric communication key established during attestation is only known by the monitor or the entity, unless they were compromised before the attestation.

More specifically, our attestation model considers an infinite number of agents and protocol executions happening in parallel. 
Since the multiset is initially empty, we model creation rules for all possible agents (e.g., clients, monitors, TEEs, CAs) to properly initialize their state and generate any long-term keys.

To translate the protocol to rules, which operate on a global state, we identify for each protocol step: \emph{(i)} the necessary inputs from the network, as well as, the persistent states for each agent, \emph{(ii)} the resulting outputs on the network, as well as, any modifications to the persistent states of each agent, \emph{(iii)} any checks performed by an agent, which may translate to restrictions of the rule transitions.
We assume the initial registration of entities on the \gdpr monitor is done over a secure channel.

To augment the attacker with capabilities to compromise long-term keys, we model rules that mark an agent as compromised and send all its secrets in plaintext on the network.

\myparagraph{Logging model}
Our logging model assumes the correct implementation of secure hardware counters and a secret key to verify log entries. We verify that two properties hold after successful log validation: \emph{(i)}~every batch stabilized before the last counter increase has a valid log entry; and \emph{(ii)}~every log entry with a counter smaller than the current stable counter belongs to a valid batch.

We model the logging layer, similarly to the attestation model, including an initialization rule that sets up the hardware counter and secret key. We model the monotonically increasing counter values symbolically and define their comparison function on the timestamps when the counter values were queried. We model the request to persist a batch, counter readout, and stabilization events as separate rules, forcing Tamarin to consider all possible orderings of these for an infinite number of batches.

The persisted log state is modeled as a series of \texttt{add} and \texttt{remove} operations; an entry is in the log if it was added but not deleted before the readout. To allow the attacker to freely modify the log, we model rules that directly create \texttt{add} or \texttt{remove} operations at any point in time. Additionally, all operations are also shared on the network to make their values accessible to the attacker.

\begin{table}[t]
\small
\centering
\begin{tabular}{l p{0.7\linewidth}}
\toprule
\textbf{Action fact} & \textbf{Description} \\
\midrule
K\((i)\) & The attacker knows information \(i\). \\
Comp\((a)\) & Actor \(a\) is compromised and its long-term keys are leaked. \\
AttE\((gm, e)\) & GDPRuler monitor \(gm\) trusts the entity \(e\). \\
AttGM\((e, gm)\) & Entity \(e\) trusts the GDPRuler monitor \(gm\). \\
ReqAtt\((e, gm)\) & Entity \(e\) initiates an attestation request for the GDPRuler monitor \(gm\). \\
SKey\((a, b, \textit{key})\) & Actor \(a\) establishes the symmetric key \(\textit{key}\) with actor \(b\) after successful attestation. \\
\bottomrule
\end{tabular}
\caption{Action facts used in the attestation model.}\label{tab:apx-verif-facts-attestation}
\vspace{-4mm}
\end{table}

\begin{table}[t]
\small
\centering
\begin{tabular}{l p{0.7\linewidth}}
\toprule
\textbf{Action fact} & \textbf{Description} \\
\midrule
LAdd\((\textit{id}, e)\) & Log entry \(e\) is added to the persisted log, and can be referenced using \(\textit{id}\). \\
LDel\((\textit{id})\) & The log entry with id \(\textit{id}\) is deleted. \\
Valid\(()\) & The persisted log is validated. \\
Store\((b)\) & Batch \(b\) is requested to be persisted. \\
Stable\((b)\) & Batch \(b\) receives confirmation that it was stabilized. \\
CtrAtt\(()\) & The hardware counter increased and attested all software counters. \\
\bottomrule
\end{tabular}
\caption{Action facts used in the logging model.}\label{tab:apx-verif-facts-logging}
\vspace{-4mm}
\end{table}

{
    \newcommand{\at}{\mathbin{\vcenter{\hbox{\text{\footnotesize{@}}}}}}

    \subsection{Verified Properties}
    To verify properties of our models, we label the rules in our models with \textit{action facts}. These action facts can be used in first-order temporal formulas. We express our desired properties by reasoning about these action facts and associated time points. We write \(\text{actionfact}(...) \at t_i\) to denote that \(\text{actionfact}(...)\) occurred at time \(t_i\). 
  
    \subsubsection{Attestation Properties}
    We verify multiple properties of the attestation model, including a set of sanity checks that ensure our model works as expected. To do this, we labeled our rules with a set of action facts, as explained in \autoref{tab:apx-verif-facts-attestation}.
    The sanity check property states that it is possible to reach the end of the protocol (successful attestation) with only a single instance of each agent and without compromising anything. Tamarin proved this property holds by finding an execution trace that upholds all these constraints. With confidence in our model established, we proved the following security properties: 

    \myparagraph{Attestation}
    We split the mutual attestation into two individual properties: (i) the monitor correctly attests any entity, and (ii) the entity correctly attests the monitor. Formally, we verify these two formulas:

    \begingroup
    \setlength{\abovedisplayskip}{-0.7em} %
    \setlength{\belowdisplayskip}{0.4em} %
    \setlength{\jot}{0.1em} %
    \begin{equation}
        \begin{split}
               \forall~gm, e, t_i: \text{AttE}(gm, e) \at t_i \\
                    \implies ( ( \exists t_j: t_j < t_i \land \text{ReqAtt}(e, gm) \at t_j ) \\
                     \lor (\exists t_c: \text{Comp}(gm) \at t_c \lor \text{Comp}(e) \at t_c ) )
        \end{split}
    \end{equation}
    \endgroup

    and 

    \begingroup
    \setlength{\abovedisplayskip}{-0.7em} %
    \setlength{\belowdisplayskip}{0.4em} %
    \setlength{\jot}{0.1em} %
    \begin{equation}
        \begin{split}
               \forall~e, gm, t_i: \text{AttGM}(e, gm) \at t_i \\
                    \implies ( ( \exists t_j: t_j < t_i \land \text{AttE}(gm, e) \at t_j ) \\
                     \lor (\exists t_c: \text{Comp}(gm) \at t_c \lor \text{Comp}(e) \at t_c ) )
        \end{split}
    \end{equation}
    \endgroup

    \myparagraph{Perfect forward secrecy (PFS)}
    We verify PFS for all symmetric session keys established after successful attestation. PFS guarantees that the session key, and thus the data encrypted with it, remains secret even if the long-term keys of either party are compromised in the future. Formally, we verify:

    \begingroup
    \setlength{\abovedisplayskip}{-0.7em} %
    \setlength{\belowdisplayskip}{0.4em} %
    \setlength{\jot}{0.1em} %
    \begin{equation}
        \begin{split}
               \forall~a, b, \textit{key}, t_i: \text{SKey}(a, b, \textit{key}) \at t_i 
                    \implies ( ( \not\exists t_k: \text{K}(\textit{key}) \at t_k ) \\
                     \lor (\exists t_c: t_c < t_i \land ( \text{Comp}(a) \at t_c \lor \text{Comp}(b) \at t_c ) ) )
        \end{split}
    \end{equation}
    \endgroup

    \begin{figure*}[t]
    \centering
    \includegraphics[]
    {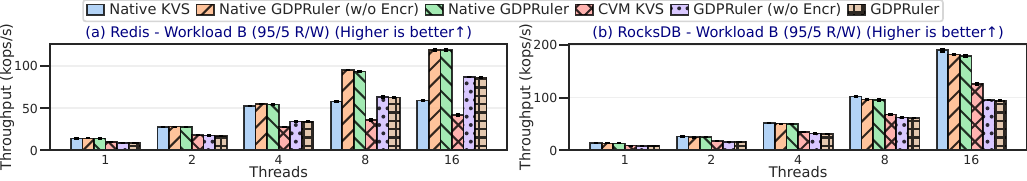}
    \vspace{-1mm}
    \caption{Scalability analysis: \projectname{}'s scalability with YCSB's workload B.}
    \vspace{-1mm}
    \label{fig:perf:scalability_b}
    \end{figure*}
    
    \begin{figure*}[t]
    \centering
    \includegraphics[]
    {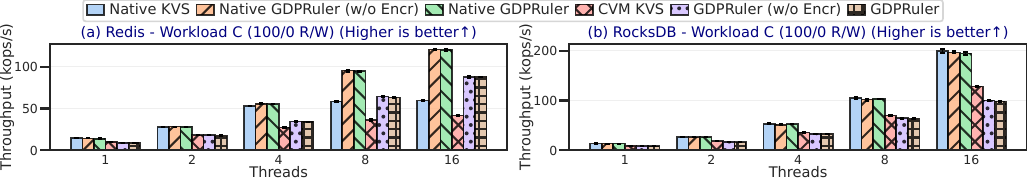}
    \vspace{-1mm}
    \caption{Scalability analysis: \projectname{}'s scalability with YCSB's workload C.}
    \vspace{-1mm}
    \label{fig:perf:scalability_c}
    \end{figure*}

    \subsubsection{Logging Properties}
    Similar to the attestation properties, we also verify a sanity property of the logging protocol to ensure our model works as expected. For the logging protocol, we ensure that it is possible to create a non-empty log that can be validated without the attacker tampering with the log. Tamarin again successfully finds a valid execution trace that upholds these constraints. 
    We verify that a validated log is complete through two properties using the action facts listed in \autoref{tab:apx-verif-facts-logging}. For readability, we use \(e(b)\) as a shorthand for the encapsulation of a batch to create a persisted log entry. Formally, we verify that the log is missing no entries:

    \begingroup
    \setlength{\abovedisplayskip}{-0.7em} %
    \setlength{\belowdisplayskip}{0.4em} %
    \setlength{\jot}{0.1em} %
    \begin{equation}
        \begin{split}
               \forall~b, t_{\textit{stable}}, t_{\textit{ctr}}, t_{\textit{valid}}: 
                    ( t_{\textit{stable}} < t_{\textit{ctr}} < t_{\textit{valid}} \\ 
                    \land \text{Stable}(b) \at t_{\textit{stable}}
                    \land \text{CtrAtt}() \at t_{\textit{ctr}}
                    \land \text{Valid}() \at t_{\textit{valid}} ) \\
                    \implies \exists \textit{id}, t_{a} : ( \text{LAdd}(\textit{id}, e(b)) \at t_{a} \\
                    \land ( \not \exists t_{d} : t_{d} < t_{\textit{valid}} \land \text{LDel}(\textit{id}) \at t_{d} ) )
        \end{split}
    \end{equation}
    \endgroup

    \noindent and that all log entries are valid:

    \begingroup
    \setlength{\abovedisplayskip}{-0.7em} %
    \setlength{\belowdisplayskip}{0.4em} %
    \setlength{\jot}{0.1em} %
    \begin{equation}
        \begin{split}
               \forall~e', \textit{id}, t_{\textit{add}}, t_{\textit{ctr}}, t_{\textit{valid}}: 
                    ( t_{\textit{add}} < t_{\textit{ctr}} < t_{\textit{valid}} \\ 
                    \land \text{LAdd}(\textit{id}, e') \at t_{\textit{add}}
                    \land \text{CtrAtt}() \at t_{\textit{ctr}}
                    \land \text{Valid}() \at t_{\textit{valid}} ) \\
                    \implies ( (\exists b, t_{s} : e' = e(b) \land t_{s} < t_{\textit{add}} \land \text{Store}(b) \at t_{s}) \\
                    \lor ( \exists t_{d} : t_{d} < t_{\textit{valid}} \land \text{LDel}(\textit{id}) \at t_{d} ) )
        \end{split}
    \end{equation}
    \endgroup
    
    \subsubsection{Verification Result}
    We successfully verified all the above properties for the attestation and logging models. Tamarin automatically verified that our models do not violate any of these properties. This automated analysis took roughly 90 seconds on an Intel(R) Xeon(R) Gold 6438Y+ processor with 500GB of RAM. 
}

\section{Additional Results}
\label{sec:appendix:additional_perf}

\begin{figure*}[t]
\centering
\includegraphics[]
{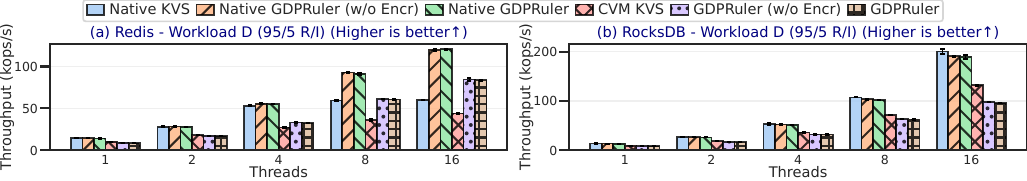}
\vspace{-1mm}
\caption{Scalability analysis: \projectname{}'s scalability with YCSB's workload D.}
\vspace{-2mm}
\label{fig:perf:scalability_d}
\end{figure*}

\begin{figure*}[t]
\centering
\includegraphics[]
{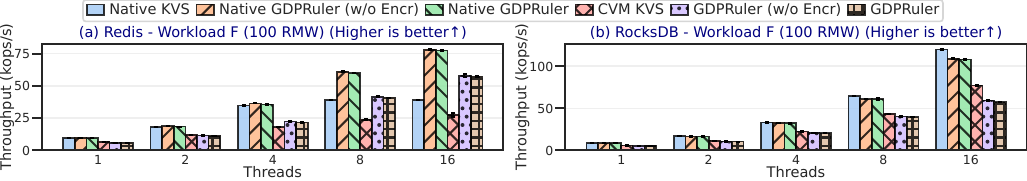}
\vspace{-1mm}
\caption{Scalability analysis: \projectname{}'s scalability with YCSB's workload F.}
\vspace{-2mm}
\label{fig:perf:scalability_f}
\end{figure*}

\subsection{Scalability Analysis}
\label{sec:appendix:scalability}

We present detailed scalability analysis for the remaining YCSB workloads (B, C, D, F) across Redis and RocksDB. We execute our experiments with varying thread counts from 1 to 16. Further, to break down \projectname{}'s overhead, we first analyze the baseline CVM cost, then examine the incremental \gdpr compliance overhead.

\myparagraph{CVM baseline overhead}
Deploying KVS engines within CVMs introduces consistent overhead due to memory encryption and I/O costs. Across all workloads, CVM KVS achieves 68-74\% of native throughput for RocksDB and 70-76\% for Redis, corresponding to 26-32\% overhead. This overhead remains stable across thread counts (±2\%), confirming that CVM costs scale proportionally with workload intensity. Importantly, the \gdpr compliance layer adds an additional 8-12\% overhead on top of this CVM baseline.

\myparagraph{Workload B (95/5 Read/Write)}
\autoref{fig:perf:scalability_b} illustrates \projectname{}'s scaling characteristics under Workload B for Redis and RocksDB. In this workload, the system exhibits similar scaling characteristics to Workload A. For RocksDB, \projectname{} maintains 58-62\% of baseline throughput across all thread counts, achieving 80.5 K ops/sec at 16 threads compared to the baseline's 129.8 K ops/sec. Breaking down the overhead: CVM introduces 27\%, while the \gdpr layer adds 9\%, totaling 36-38\% overhead. \projectname{} with Redis shows comparable behavior at lower thread counts (64-66\% of baseline) but benefits from Unix socket advantages beyond 8 threads, reaching 89.2 K ops/sec versus the baseline's 71.4 K ops/sec at 16 threads—a 25\% improvement demonstrating that \projectname{}'s connection handling overcomes Redis's server TCP bottleneck.

\myparagraph{Workload C (100/0 Read/Write)}
\projectname{}'s scalability in the read-only Workload C is shown in \autoref{fig:perf:scalability_c} for both KVS backends.
As the most read-intensive workload, Workload C achieves higher absolute throughputs while maintaining similar relative overheads. \projectname{} with RocksDB delivers 97.3 K ops/sec at 16 threads (49.4\% of 197.0 K ops/sec baseline), where CVM accounts for 28\% overhead and \gdpr compliance adds 11\%. \projectname{} with Redis achieves 101.0 K ops/sec (119\% of 84.7 K ops/sec baseline), with the crossover point occurring at 8 threads. Both databases demonstrate near-linear scaling efficiency (75-80\%) up to 16 threads, with \projectname{}'s overhead remaining constant across scaling, indicating no additional synchronization bottlenecks.

\myparagraph{Workload D (95/5 Read/Insert)}
\autoref{fig:perf:scalability_d} depicts \projectname{}'s scalability results for Workload D on Redis and RocksDB.
Workload D, emphasizing recent data access patterns, shows performance characteristics between Workloads A and C. \projectname{} achieves 56-60\% of baseline performance for RocksDB, with CVM contributing 26\% overhead and \gdpr adding 10\% on top of it. At 16 threads, \projectname{} with RocksDB reaches 76.8 K ops/sec (58.1\% of 132.2 K ops/sec), while with Redis it delivers 95.4 K ops/sec (128\% of 74.3 K ops/sec). The stable 9-11\% \gdpr layer overhead across all thread counts compared to the CVM KVS validates that metadata processing scales efficiently without introducing contention.

\myparagraph{Workload F (100\% Read-Modify-Write)}
\autoref{fig:perf:scalability_f} presents the scalability results for \projectname{}'s experiments with Workload F both for Redis and RocksDB.
Workload F exercises read-modify-write operations. \projectname{} maintains consistent performance ratios: 59-63\% for RocksDB and improved relative performance for Redis at higher thread counts. \projectname{} with RocksDB achieves 68.9 K ops/sec at 16 threads (60.2\% of 114.5 K ops/sec), with CVM overhead at 29\% and \gdpr at 8\% (compared to CVM KVS). \projectname{} with Redis backend reaches 78.3 K ops/sec (131\% of 59.7 K ops/sec). The uniform 8-10\% \gdpr layer overhead across this workload confirms that transactional metadata handling does not introduce additional synchronization costs beyond the base CVM overhead.

\begin{figure}[t]
\centering
\includegraphics[]
{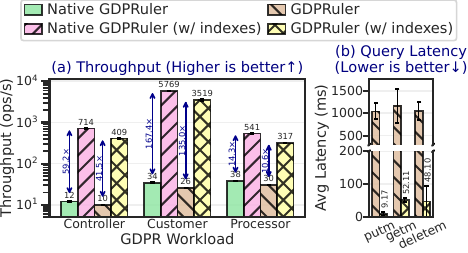}
\vspace{-1mm}
\caption{\projectname{}'s performance with YCSB workloads for RocksDB backend.}
\vspace{-1mm}
\label{fig:gdpr_metadata_perf_rocksdb}
\end{figure}

\myparagraph{RocksDB \gdpr workload performance}
RocksDB exhibits similar indexing benefits to Redis while achieving higher absolute throughput, as shown in~\autoref{fig:gdpr_metadata_perf_rocksdb}. Without indexes, \projectname{} must perform full KVS scans, resulting in performance as low as 9.9, 26.0, and 29.8 ops/sec for Controller, Customer, and Processor workloads respectively. When metadata indexes are enabled, performance improves dramatically by 10.6-135$\times$ depending on the workload. Customer workloads benefit most from indexed ownership lookups (135$\times$ speedup), while Processor workloads achieve 10.6$\times$ faster execution through purpose-based filtering. These improvements translate to absolute throughputs of 408.8, 3,518.9, and 316.8 ops/sec for the three workloads in CVM deployments.

\gdpr-operation-level latencies for RocksDB further confirm the effectiveness of indexing: \emph{getm} operations improve from 1,164 ms to 52 ms (22.3$\times$ speedup), while \emph{putm} operations decrease from 1,041 ms to 9 ms (113$\times$ speedup). Standard operations (\emph{get}, \emph{put}, \emph{delete}) also benefit from reduced contention, showing 21-33\% latency improvements despite not leveraging metadata indexes. The CVM deployment introduces 39-43\% overhead compared to native execution with indexes enabled, consistent with the CVM baseline costs observed in YCSB workloads. These results validate that metadata indexing is critical for \gdpr compliance queries regardless of the underlying KVS engine, and that the indexing benefits far outweigh confidential computing costs.
\begin{table}[H]
\centering
\footnotesize
\begin{tabular}{@{}cccccc@{}}
\toprule
\multirow{2}{*}{\parbox{0.8cm}{\centering\textbf{Entry\\ size}}} & \multirow{2}{*}{\parbox{0.8cm}{\centering\textbf{Data\\ size}}} & \multicolumn{4}{c}{\textbf{Compression ratio (\% data size reduction)}} \\
\cmidrule(l){3-6}
 & & Level 0 & Level 3 & Level 6 & Level 9 \\
\midrule
1KB & 10GB & 0.99 (--0.6\%) & 2.06 (51.3\%) & 2.13 (53.1\%) & 2.15 (53.5\%) \\
\addlinespace
4KB & 10GB & 1.00 (--0.2\%) & 2.07 (51.8\%) & 2.17 (53.9\%) & 2.20 (54.6\%) \\
\bottomrule
\end{tabular}
\caption{Compression effectiveness across entry sizes.}
\vspace{-2mm}
\label{tab:microbenchmark-compression}
\end{table}

\begin{table*}[t]
\centering
\small
\caption{Client rights and controller obligations in GDPR, CCPA, and VCDPA that are relevant for data-management systems.}
\label{tab:privacy_laws_comparison}
\vspace{-0.1cm}
\begin{tabular}{|p{3cm}|p{4.2cm}|p{4.2cm}|p{4.2cm}|}
\hline
\textbf{Obligation/Right} & \textbf{GDPR (EU)} & \textbf{CCPA (California)} & \textbf{VCDPA (Virginia)} \\
\hline
\multicolumn{4}{|c|}{\textbf{Client Rights}} \\
\hline
Right to Access & Article 15~\cite{gdpr-art-15} -- Full access to personal data held & $\S$ 1798.110~\cite{ccpa-sec-1798.110}, $\S$ 1798.115~\cite{ccpa-sec-1798.115} -- Access to categories, data, and sharing information & $\S$ 59.1-577(A)(1)~\cite{vcdpa-sec-59.1-577} -- Confirm processing and access personal data \\
\hline
Right to Deletion & Article 17~\cite{gdpr-art-17-leg} -- Right to be Forgotten (broad grounds) & $\S$ 1798.105~\cite{ccpa-sec-1798.105} -- Delete collected data & $\S$ 59.1-577(A)(3)~\cite{vcdpa-sec-59.1-577} -- Delete personal data provided or obtained \\
\hline
Right to Portability & Article 20~\cite{gdpr-art-20} -- Data in a common, machine-readable format and right to transmit & $\S$ 1798.130~\cite{ccpa-sec-1798.130} -- Data transmission between organizations and data subjects & $\S$ 59.1-577(A)(4)~\cite{vcdpa-sec-59.1-577} -- Copy of data in portable, usable format that allows for transmission to another controller\\
\hline
Right to Object/Opt-Out & Article 21~\cite{gdpr-art-21} -- Object to processing (e.g., marketing) & $\S$ 1798.120~\cite{ccpa-sec-1798.120} -- Opt-out of ``sale'' and ``sharing'' (cross-context ads) & $\S$ 59.1-577(A)(5)~\cite{vcdpa-sec-59.1-577} -- Opt-out of targeted advertising, sale, and profiling \\
\hline
Right to Rectification & Article 16~\cite{gdpr-art-16} -- Correct inaccurate data & $\S$ 1798.106~\cite{ccpa-sec-1798.106} -- Correct inaccurate personal information & $\S$ 59.1-577(A)(2)~\cite{vcdpa-sec-59.1-577} -- Correct inaccuracies in personal data \\
\hline
Right to Restrict / Limit Processing & Article 18~\cite{gdpr-art-18-leg} -- Restrict processing of data for specified purposes & $\S$ 1798.121~\cite{ccpa-sec-1798.121} -- Limit use of \textit{Sensitive} Personal Information  & Not explicitly provided (but consent required for sensitive data $\S$ 59.1-578~\cite{vcdpa-sec-59.1-578}) \\
\hline
Automated Individual Decision-Making & Article 22~\cite{gdpr-art-22} -- Right not to be subject to solely automated decisions & $\S$ 1798.185(a)(15)~\cite{ccpa-sec-1798.185} -- No GDPR-style statutory right; rulemaking authority to regulate automated decision making technologies & $\S$ 59.1-577(A)(5)~\cite{vcdpa-sec-59.1-577} -- Opt-out of profiling for legal/significant effects \\
\hline
\multicolumn{4}{|c|}{\textbf{Controller Obligations}} \\
\hline
Legal Basis for Processing & Article 6~\cite{gdpr-art-6} -- Explicit basis required (consent, contract, legal obligation, legitimate interest, etc.) & $\S$ 1798.100~\cite{ccpa-sec-1798.100}, $\S$ 1798.115~\cite{ccpa-sec-1798.115}, $\S$ 1798.120~\cite{ccpa-sec-1798.120} -- Must provide notice on collection \& opt-out option & $\S$ 59.1-578~\cite{vcdpa-sec-59.1-578} -- Processing limits based on necessity/purpose; Consent for sensitive data and associated privacy notice\\
\hline
Security \& Risk Assessments & Article 32~\cite{gdpr-art-32}, 35~\cite{gdpr-art-35} -- Technical measures \& Data Protection Impact Assessment mandatory for high risk & $\S$ 1798.185~\cite{ccpa-sec-1798.185} -- May require annual cybersecurity audits and risk assessments for high risk & $\S$ 59.1-580~\cite{vcdpa-sec-59.1-580} -- Data protection assessments required for targeted ads, sale, profiling \\
\hline
Audit Logging / Accountability & Article 30~\cite{gdpr-art-30}, 33~\cite{gdpr-art-33}, 34~\cite{gdpr-art-34}, 5(1)(f)~\cite{gdpr-art-5-leg} -- Detailed logging of access, modification, consent, and breach notifications mandatory & Implicit in $\S$ 1798.185~\cite{ccpa-sec-1798.185} (Audits) and in $\S$ 1798.105(c)~\cite{ccpa-sec-1798.105} (Confidential record of deletion requests) & $\S$ 59.1-580(D)~\cite{vcdpa-sec-59.1-580}, $\S$ 59.1-578~\cite{vcdpa-sec-59.1-578} -- Controller must comply and demonstrate compliance \\
\hline
Breach Notification & Article 33~\cite{gdpr-art-33} -- Notify authority within 72 hours & $\S$ 1798.82~\cite{ccpa-sec-1798.82} -- Notify ``without unreasonable delay'' & Va. Code $\S$ 18.2-186.6~\cite{va-code-18.2-186.6} -- Notify ``without unreasonable delay'' \\
\hline
Privacy Notice Requirements & Article 13~\cite{gdpr-art-13}, 14~\cite{gdpr-art-14} -- Comprehensive transparency required & $\S$ 1798.100~\cite{ccpa-sec-1798.100} -- Notice at collection; 12-month look-back on sharing & $\S$ 59.1-578(C)~\cite{vcdpa-sec-59.1-578} -- Reasonably accessible, clear, and meaningful privacy notice \\
\hline
Encryption / Data Protection / Data Security & Article 5(1)(f)~\cite{gdpr-art-5-leg}, 25~\cite{gdpr-art-25} -- Appropriate security measures for integrity and confidentiality & $\S$ 1798.100(e)~\cite{ccpa-sec-1798.100}, $\S$ 1798.81.5~\cite{ccpa-sec-1798.81.5} -- Reasonable security procedures and practices appropriate to the nature of the personal information & $\S$ 59.1-578(A)(3,4)~\cite{vcdpa-sec-59.1-578} -- Administrative, technical, and physical data security practices \\
\hline
Data Retention Limits & Article 5(1)(e)~\cite{gdpr-art-5-leg} -- Storage limitation & $\S$ 1798.100(a)(3)~\cite{ccpa-sec-1798.100} -- Not longer than reasonably necessary for disclosed purpose & $\S$ 59.1-578~\cite{vcdpa-sec-59.1-578}, $\S$ 59.1-579(2)~\cite{vcdpa-sec-59.1-579} -- Not explicitly capped (at the controller's discretion), but limited to what is relevant/necessary \\
\hline
\end{tabular}
\end{table*}

\subsection{Compression}

To validate the effectiveness of the utilized compression layer, we perform additional microbenchmarks using realistic \gdpr log entries containing semi-compressible payloads with common patterns (JSON structures, repeated field names, and random data). Table~\ref{tab:microbenchmark-compression} shows results from processing 10GB of log data across different entry sizes (1KB, 4KB) and compression levels (0, 3, 6, 9). \projectname{}'s compression system achieves substantial storage reduction, with compression levels 3-9 providing 51-55\% space savings. Larger entries (4KB) exhibit slightly better compression ratios (2.20:1 at level 9) due to more repeated patterns within entries.
Compression level 3 provides 99\% of the space savings achieved by level 9 with significantly lower CPU consumption, making it \projectname{}'s default configuration choice.

\section{Legislation framework analysis}
\label{sec:appendix:legislation}
To highlight the versatility and applicability of \projectname{} in other regulatory framework, we present Table~\ref{tab:privacy_laws_comparison} (next page), which summarizes the subset of client rights and controller obligations in General Data Protection Regulation (\gdpr), California Consumer Privacy Act (CCPA), and Virginia’s Consumer Data Protection Act (VCDPA) that are directly actionable at the data-management layer and thus relevant to \projectname{}'s design. The table highlights the shared structural obligations, such as rights to access, deletion, portability, objection/opt-out, and duties around security, logging, and data retention. The observed alignment supports a regulation-agnostic view of \projectname{}'s policy language: the same metadata fields (e.g., ownership, retention times, purposes, objection flags) and enforcement hooks can be instantiated under different ``profiles'' (e.g., GDPR, CCPA, VCDPA), with regulation-specific nuances captured by the choice of predicates and enforcement rules rather than by redesigning the entire underlying system.

}
{

}

\end{document}